\documentclass[11pt,a4paper]{article}
\pdfoutput=1
\usepackage{jcappub}

\usepackage{amsmath}
\usepackage{amssymb}
\usepackage{graphicx}
\usepackage{color}

\usepackage{subfigure}

\newcommand{\tens}[1]{\mbox{\bf \sf{#1}}}

\renewcommand{\vec}[1]{{\bf #1}}

\newcommand{\mtrx}[1]{\ensuremath{\tens{{#1}}}}
\newcommand{\prob}{\ensuremath{{\rm P}}}
\newcommand{\expectation}{\ensuremath{\mathbb{E}}}
\newcommand{\lmax}{\ell_{\rm max}}
\newcommand{\nside}{N_{\rm side}}
\newcommand{\alm}{a_{\ell m}}

\allowdisplaybreaks

\title{Sparse Inpainting and Isotropy}

\author[a]{Stephen M. Feeney,}
\author[b]{Domenico Marinucci,}
\author[a,c]{Jason D. McEwen,}
\author[a,d]{Hiranya V. Peiris}
\author[d,e,f,g]{Benjamin D. Wandelt}
\author[b]{and Valentina Cammarota}
\emailAdd{s.feeney@imperial.ac.uk}

\affiliation[a]{Department of Physics and Astronomy, University College London, London WC1E 6BT, UK}
\affiliation[b]{Department of Mathematics, University of Rome Tor Vergata, via della Ricerca Scientifica 1, 00133 Roma, Italy}
\affiliation[c]{Mullard Space Science Laboratory (MSSL), University College London, Surrey RH5 6NT, UK}
\affiliation[d]{Kavli Institute for Theoretical Physics, Kohn Hall, University of California, Santa Barbara, CA 93106, USA}
\affiliation[e]{Institut d'Astrophysique de Paris, UMR 7095, CNRS - Universit\'e Pierre et Marie Curie (Univ Paris 06), 98 bis blvd Arago, 75014 Paris, France}
\affiliation[f]{Sorbonne Universit\'e, Institut Lagrange de Paris (ILP), 98 bis bd Arago, 75014 Paris, France}
\affiliation[g]{Departments of Physics and Astronomy, University of Illinois at Urbana-Champaign, Urbana, IL 61801, USA}

   \abstract{Sparse inpainting techniques are gaining in popularity as
     a tool for cosmological data analysis, in particular for handling
     data which present masked regions and missing observations. We
     investigate here the relationship between sparse 
     inpainting techniques using the spherical harmonic basis as a dictionary 
     and the isotropy properties of cosmological maps, as for instance
     those arising from cosmic microwave background (CMB)
     experiments. In particular, we investigate the possibility that
     inpainted maps may exhibit anisotropies in the behaviour of
     higher-order angular polyspectra. We provide analytic
     computations and simulations of inpainted maps for a Gaussian
     isotropic model of CMB data, suggesting that the resulting
     angular trispectrum may exhibit small but non-negligible
     deviations from isotropy.}

\begin{document}
\maketitle


\section{Introduction}

Sparse inpainting techniques are becoming widespread tools for cosmic
microwave background (CMB) data analysis. The reasons for their
popularity are easily understood: on one hand these techniques belong
to the rich framework of convex optimization procedures, and are
related to deep and elegant mathematical results on compressive
sensing (e.g., refs.~\cite{donoho:2006, donoho:2006b, baraniuk:2010,
  candes:2006b, rauhut:2011, rauhut:2012}); on the other hand they
address some important issues in practical cosmological data analysis,
e.g., dealing with masked regions and missing observations. Consequently, 
inpainting techniques on the sphere \citep{abrial:2007,
  rauhut:2011, rauhut:2012, starck:2013b, starck:2013a, mcewen:css2}
have now found successful applications for many astrophysical problems
(see, e.g., refs.~\cite{abrial:2007,dupe:2010,starck:2010} and references
therein).

Given the utility of these approaches for modern cosmological data
analysis, it is important to investigate their properties from many
different points of view. In particular, we shall be concerned here
with the relationship between inpainting techniques and isotropy: we
shall investigate whether inpainted maps retain the same isotropy
properties as the input random fields from which they are derived. We
shall show that, in principle, inpainting can produce anisotropies in
CMB-like maps, in the form of a small modulation of the angular
trispectrum. For brevity's sake, we consider only the direct
application of sparse inpainting techniques to low-$\ell$ CMB-only
maps. Although extra factors, such as instrumental noise and beams and
variable sky coverage, are present in more realistic scenarios, we
conjecture that they will not substantially affect the core
conclusions of this paper; we leave further investigation on this
point to future research.

Our basic arguments can be summarized as follows. To understand
the relationship between inpainting and isotropy, it can be
convenient to recognize sparse inpainting with a spherical harmonic dictionary as operationally
equivalent to maximizing a Bayesian posterior distribution
assuming a Laplacian prior on the spherical harmonic coefficients
$a_{\ell m}$. Earlier results from ref.~\cite{baldi:2006} (see also
refs.~\cite{baldi:2007,marinucci:2011,baldi:2013}) have shown that a
random field generated by sampling such independent non-Gaussian
coefficients is necessarily anisotropic. We demonstrate that the
trispectrum of such a field exhibits a modulated behavior,
containing a maximum at the North Pole and oscillations whose
pattern can be analytically derived. Of course, the distribution
of maximum \emph{a posteriori} estimates is, in general, different from
the law of the input prior; it is natural, however, to conjecture that the
trispectrum of inpainted maps will at least partially inherit
these anisotropic features. This conjecture is indeed confirmed by
our simulations, where CMB-like Gaussian isotropic maps are
generated at low resolution, masked, and then inpainted: their
trispectra contain a similar pattern to that analytically derived from 
the theoretical considerations, although with a smaller absolute
amplitude.\footnote{The Fortran code used to produce and 
inpaint these simulations is available for download from 
\url{http://zuserver2.star.ucl.ac.uk/~smf/code.html}.} 
It should be noted that our arguments and 
results apply only to sparse inpainting using the spherical 
harmonics as a dictionary, as proposed in ref.~\cite{starck:2013b}; other dictionary choices, such as wavelets, 
may be free from such issues.

As was previously mentioned, the simulations are performed in
somewhat idealized circumstances -- namely at low band-limit
(maximum harmonic mode \mbox{$\ell_{\mathrm{max}}=10$}) and in the
absence of noise or an instrumental beam -- to ensure speed of
calculation and clarity of results. A large sky cut is used to
demonstrate the main result; a more moderate sky cut has also been
shown to produce similar effects. Further investigations under
more realistic conditions are left as an avenue for further
research. We note that the trispectrum modulations for $\ell=2$
and $\ell=3$ exhibit qualitatively the same pattern; this could
produce a spurious correlation between multipoles which are
actually independent.

It is important to stress that we do not, by any means, view this paper
as part of the long-standing debate between Bayesian and
frequentist approaches in astrophysical data analysis; the authors
of this paper have rather different points of view on these
foundational issues. The Bayesian interpretation of inpainting
techniques is introduced here merely as a heuristic rationale to
motivate the investigation of the expected sample properties of
inpainted maps: these properties are then considered in a setting
which can be viewed as purely frequentist, i.e., by deriving
analytically, and then supporting by simulations, the expected
modifications of a standard trispectrum estimator defined on the
inpainted maps.  It would be possible, in principle, to correct
the trispectrum modifications that result, but this is likely to
be very demanding from the computational point of view.

The structure of the paper is as follows: in
Sect.~\ref{sec:inpainting}, we review convex regularization
techniques and their relationship to Laplacian priors; in
Sect.~\ref{sec:gaussianity} we review earlier results from
refs.~\cite{baldi:2006,baldi:2007} on the relationship
between Gaussianity, independent Fourier coefficients and
isotropy; in Sect.~\ref{sec:results} we present our simulations to
support the findings of Sect.~\ref{sec:gaussianity}; concluding
remarks are made in Sect.~\ref{sec:conclusions}. Most analytic
computations are collected in Appendix~\ref{sec:appendix}.


\section{Sparse Inpainting}
\label{sec:inpainting}

It is well-known that sparse inpainting can be viewed as maximum
\emph{a posteriori} (MAP) estimation under a Laplacian prior
(e.g., ref.~\cite{gribonval:2011}).  In this section, we make the
association explicit in the context of signals defined on the
sphere, while also making a comparison to Wiener filtering.

Let us model the observed spatial data $\vec{d}$ by
\begin{equation*}
\vec{d}=\mtrx{Y}\vec{a}+\vec{n}\text{ ,}
\end{equation*}%
where $\mtrx{Y}$ is the matrix with columns given by the spherical
harmonic functions $Y_{\ell m}$ evaluated at the locations of the
observed data, $\vec{a}$ is the vector of associated
harmonic coefficients $a_{\ell m}$, and $\vec{n}$ is the noise on the
measured spatial data.  Assuming Gaussian noise with covariance
$\mtrx{N} = \expectation( \vec{n} \vec{n}^\dagger )$, where
$\expectation(\cdot)$ denotes expectation and $\cdot^\dagger$ denotes
the Hermitian transpose, the likelihood of the data is
\begin{equation*}
\prob(\vec{d} \, | \, \vec{a})
\propto \mathrm{exp}\bigl(-\chi ^{2}/2\bigr),
\end{equation*}%
where%
\begin{equation*}
\chi ^{2}=(\vec{d}-\mtrx{Y}\vec{a})^{\dagger}\mtrx{N}^{-1}(\vec{d}-\mtrx{Y}\vec{a})\text{ }.
\end{equation*}%
The posterior distribution of the harmonic coefficients $\vec{a}$ is
then given by%
\begin{equation*}
\prob(\vec{a} \, | \, \vec{d}) \propto \prob(\vec{d} \, | \, \vec{a})
\, \prob(\vec{a}).
\end{equation*}

We consider MAP estimates of the $a_{\ell
  m}$s.\footnote{Alternatively, sampling methods can be used to
  recover the full posterior distribution $\prob( \vec{a} \, | \,
  \vec{d})$ (e.g. refs.~\cite{wandelt:2004,taylor:2008}).}  Firstly, assume the
Gaussian prior distribution on the harmonic coefficients
\begin{equation*}
\prob_{\mathrm{Gauss}}(\vec{a}) \propto
\mathrm{exp} \bigl(-\vec{a}^{\dagger}\mtrx{C}^{-1}\vec{a}/2 \bigr)\text{ },
\end{equation*}%
with covariance $\mtrx{C} = \expectation( \vec{a} \vec{a}^\dagger
)$. This is a well-motivated physical prior for the CMB anisotropies,
since it is well-known that, to first approximation, the primordial
perturbations from inflation that are thought to have sourced these
anisotropies are Gaussian.  The MAP estimator under this prior
corresponds to the usual Wiener filtering approach to inpainting
(e.g., refs.~\cite{tegmark:1997,feeney:2011c,elsner:2012,elsner:2013}),
i.e., recovering harmonic components from incomplete sky data, where
the $a_{\ell m}$s are given by the solution to the optimization
problem
\begin{equation*}
\hat{\vec{a}}_{\mathrm{Wiener}}=
\underset{\vec{a}}{\arg \min} \:
 \bigl ( \vec{a}^{\dagger}\mtrx{C}^{-1}\vec{a}+\chi ^{2} \bigr)
\text{ },
\end{equation*}
where the minimization is with respect to $\vec{a}$.
This optimization problem can be solved analytically to give
\begin{equation*}
\hat{\vec{a}}_{\mathrm{Wiener}}
= (\mtrx{Y}^\dagger\mtrx{N}^{-1}\mtrx{Y}+\mtrx{C}^{-1})^{-1}
\mtrx{Y}^{\dagger}\mtrx{N}^{-1} \vec{d}\text{ }.
\end{equation*}%
Secondly, consider an alternative prior where each $a_{\ell m}$ is
independent and Laplacian-distributed:%
\begin{equation*}
\prob(\vec{a})=\prod_{\ell m} \prob_{\mathrm{Laplacian}}(a_{\ell
m})
\propto
\mathrm{exp}\bigl(-\beta \, \| \vec{a} \|_{1} \bigr)
\end{equation*}%
for scale parameter $\beta$, where we recall that the $\ell_1$ norm
of a vector is given by the sum of the absolute values of its entries,
i.e., $\| \vec{a} \|_{1} = \sum_{\ell m}|a_{\ell m}|$.  The MAP
estimator corresponding to this prior is thus recovered by solving the
optimization problem
\begin{equation*}
\hat{\vec{a}}_{\mathrm{Laplacian}}=
\underset{\vec{a}}{\arg \min} \:
 \bigl ( \beta \, \| \vec{a} \|_1 + \chi ^{2}/2 \bigr)
\text{ }.
\end{equation*}%
For the case of isotropic white noise, $\mtrx{N} = \sigma^2
\mathbb{I}$, where $\mathbb{I}$ is the identity, the optimization
problem may be rewritten
\begin{equation}
\label{eqn:optimization_problem_1}
\hat{\vec{a}}_{\mathrm{Laplacian}}=
\underset{\vec{a}}{\arg \min} \:
 \bigl ( \lambda \, \| \vec{a} \|_1 + \| \vec{d}-\mtrx{Y}\vec{a} \|_2^2 \bigr)
\text{ },
\end{equation}
for $\lambda= 2 \beta \sigma^2 >0$, which acts as a regularization 
parameter, trading off sparsity against data fidelity. This is exactly the optimization 
problem described in ref.~\cite{starck:2013a}, eqns. (4-6), and is known as an 
{\em unconstrained} optimization, as a single object function (given by 
the sum of two terms) is optimized in the absence of any additional 
constraints.

Alternatively, {\em constrained} problems are sometimes considered, e.g.,
\begin{equation}
\label{eqn:optimization_problem_2} \hat{\vec{a}}=
\underset{\vec{a}}{\arg \min} \: \| \vec{a} \|_1 \text{ such that
} \vec{d} = \mtrx{Y}\vec{a} \,,
\end{equation}
avoiding the need to choose a regularization parameter $\lambda$.%
\footnote{An alternative constrained problem can also be considered:
\begin{equation}
\hat{\vec{a}}=
\underset{\vec{a}}{\arg \min} \: \| \vec{a} \|_1 \text{ such that
} \| \vec{d}-\mtrx{Y}\vec{a} \|_2^2 \leq \epsilon^2 \text{ },
\end{equation}
where $\epsilon$ is chosen from the statistical properties of
the noise in the observations (e.g., ref.~\cite{mcewen:css2}).}
This is the problem solved by the sparse inpainting algorithm
described in Appendix B of ref.~\cite{starck:2013b} and
implemented here.
The unconstrained and constrained problems are subtly different, 
but the unconstrained problem can nevertheless be used to gain 
insight (via its Bayesian interpretation) into the constrained problem
(indeed, this intuition is born out in the analytic and simulated results 
presented subsequently).

We remark that viewing sparse inpainting as MAP estimation under a
Laplacian prior is only one possible Bayesian interpretation. It
has been shown by ref.~\cite{gribonval:2011} that, for white Gaussian
noise, the Bayesian minimum mean-square error (MMSE) estimator for
any prior is a solution to the penalized
least-squares optimization problem with a suitable penalty
function.  Nevertheless, viewing sparse inpainting as MAP estimation 
under a Laplacian prior is indeed a valid interpretation; 
furthermore, in the case of a Gaussian prior, the MAP and MMSE 
estimators are equivalent.

We also remark that it is possible to consider two alternative
implementation schemes for the optimization problems in
eqns. (\ref{eqn:optimization_problem_1}) and 
(\ref{eqn:optimization_problem_2}), according to the different
definitions we can take for the $\ell_1$ norm.  In particular, one
can either define
\begin{equation}
\label{eqn:l1norm1} \| \vec{a} \|_1 = \sum_{\ell m} | \Re a_{\ell m} | +
| \Im a_{\ell m} |
\end{equation}
or
\begin{equation}
\label{eqn:l1norm2} \| \vec{a} \|_1 = \sum_{\ell m} \sqrt{(\Re a_{\ell
m})^2 + (\Im a_{\ell m})^2} \, ;
\end{equation}
the former is the $\ell_1$ norm of the separate real and imaginary 
components of the complex $\alm$s, whereas the latter is the $\ell_1$ 
norm of the magnitude of the complex $\alm$s.
We therefore refer to inpainting based on eqn.~(\ref{eqn:l1norm1}) as
\emph{separate} sparse inpainting, and that based on
eqn.~(\ref{eqn:l1norm2}) as \emph{joint} sparse inpainting. The
analytic computations in Appendix~\ref{sec:appendix} are based on
eqn.~(\ref{eqn:l1norm1}), while the simulations cover both cases.

The constrained problem, eqn. (\ref{eqn:optimization_problem_2}), can be 
solved by an iterative algorithm based on 
Douglas-Rachford splitting~\citep{combettes:2007,combettes:2009,starck:2013b}. Each 
iteration of this algorithm consists of two steps. In the first step, the data 
constraint is applied to the latest estimate of the $\alm$s, denoted 
$\vec{a}^n$, using a projection operator:
\begin{equation}
\vec{a}^{n+\frac{1}{2}} = {\rm Proj}_{\{\vec{a}:\vec{d}=\mtrx{Y}\vec{a}\}}(\vec{a}^n).
\label{eqn:dr_proj_step}
\end{equation}
The projection operator creates a map from the $\vec{a}^n$, sets the 
unmasked pixels of this map to their observed values, $\vec{d}$, and 
returns the spherical harmonic coefficients of this constrained map, 
$\vec{a}^{n+\frac{1}{2}}$. In the second step of each iteration, the sparsity 
prior is applied to the constrained harmonic coefficients. Mathematically, this
operation takes the form
\begin{equation}
\vec{a}^{n+1} = \vec{a}^n + \Big(
{\rm sgn} \big( 2\vec{a}^{n+\frac{1}{2}} - \vec{a}^n \big) \,
{\rm max} \big(0, \, |2\vec{a}^{n+\frac{1}{2}} - \vec{a}^n| - \lambda \big) \Big)
 - \vec{a}^{n+\frac{1}{2}} ,
\label{eqn:dr_prox_step}
\end{equation}
where $\lambda > 0$ corresponds again to a weight applied to 
$\| \vec{a} \|_1$ and governs the rate of convergence of the 
algorithm\footnote{``Rate'' here is taken to mean the overall computation 
speed, not the asymptotic complexity of the algorithm (e.g., 
$\mathcal{O}(n^x)$). Changing $\lambda$ changes the 
constant of proportionality relating the complexity to the computation speed.} 
but, in theory, does not affect the solution.
Roughly speaking, this operation sets the constrained $\alm$s with 
magnitudes less than $\lambda$ to zero, and reduces the magnitudes 
of the others by $\lambda$, thus promoting a sparse solution. 
We choose $\lambda$ to be unity, and verify using both simulations with
varying $\lambda$ and convergence checks (discussed in further detail 
in Sect.~\ref{sec:results}) that the precise value chosen is unimportant. We
initialize the algorithm with a warm start to speed up convergence, taking the 
cut-sky $\alm$s as the starting point, $\vec{a}^0$; 
again, we verify through simulations that the precise choice of $\vec{a}^0$ 
does not affect the results.

The sequence of $\alm$s produced by the {\em first} step, eqn. 
(\ref{eqn:dr_proj_step}), converges to a global minimizer of eqn. (\ref{eqn:optimization_problem_2})~\citep{combettes:2007}. Applying this algorithm as written
(i.e., to vectors of complex spherical harmonic coefficients) therefore implements 
joint sparse inpainting; as eqn. (\ref{eqn:l1norm1}) is equivalent to the 
$\ell_1$ norm of the {\em real} $\alm$s, we can perform separate sparse 
inpainting by replacing the complex $\alm$s in eqns. 
(\ref{eqn:dr_proj_step}) and (\ref{eqn:dr_prox_step}) with their real 
counterparts.

\begin{figure}
\centering
  \subfigure[input map]{\includegraphics[width=0.49\textwidth]{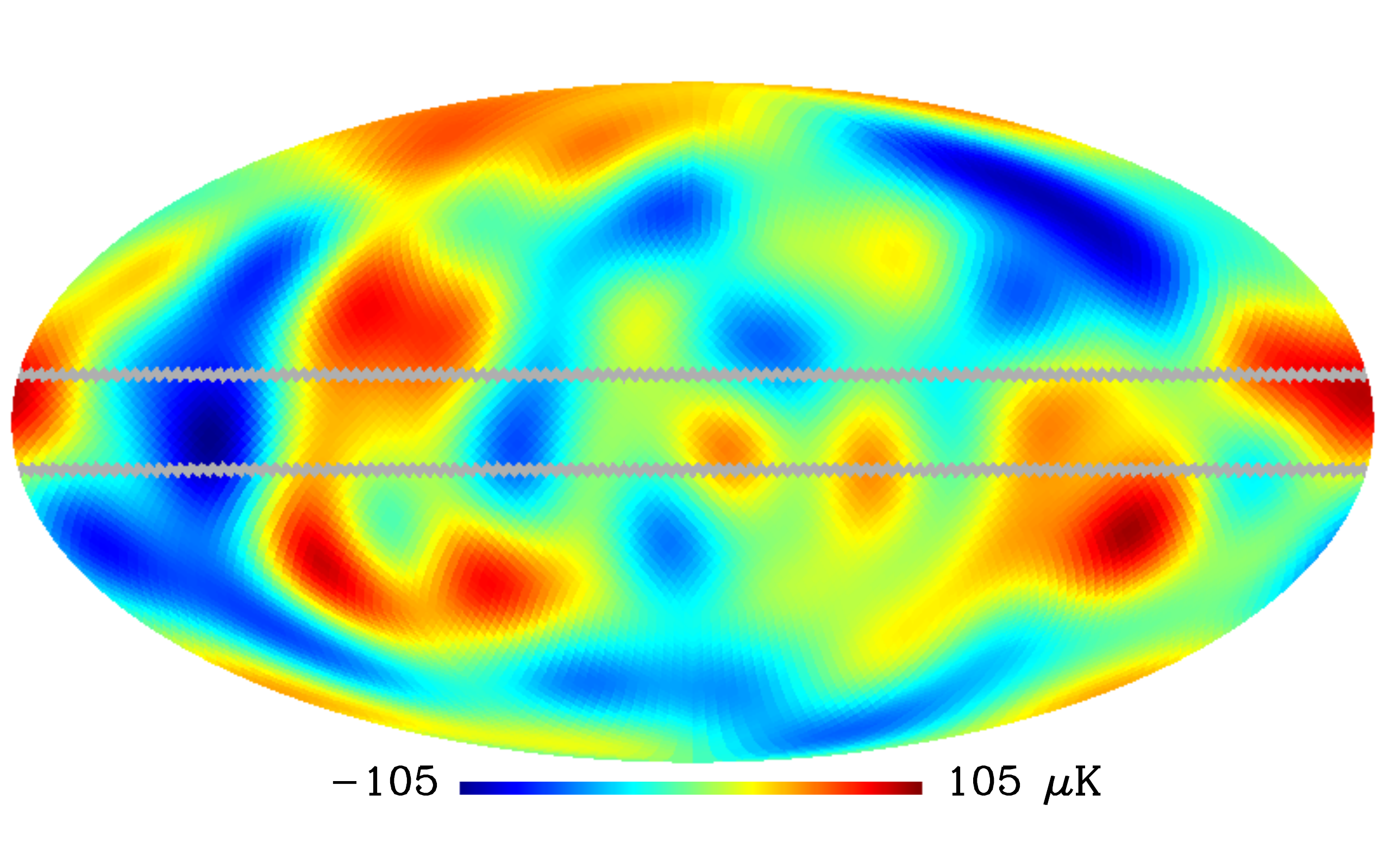}} \hfill
  \subfigure[norm evolution]{\includegraphics[width=0.49\textwidth]{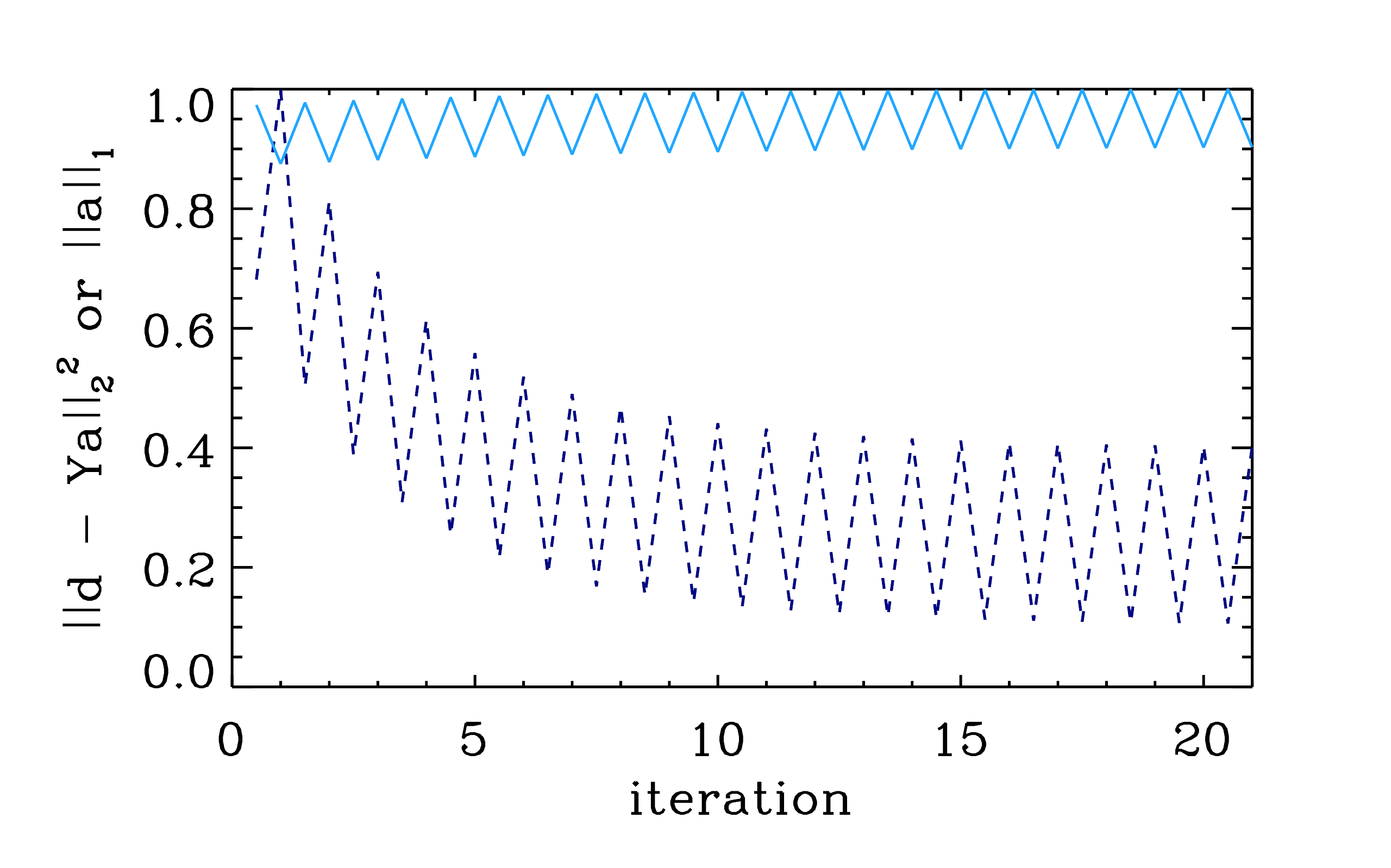}}
  \subfigure[final data-constraint step]{\includegraphics[width=0.49\textwidth]{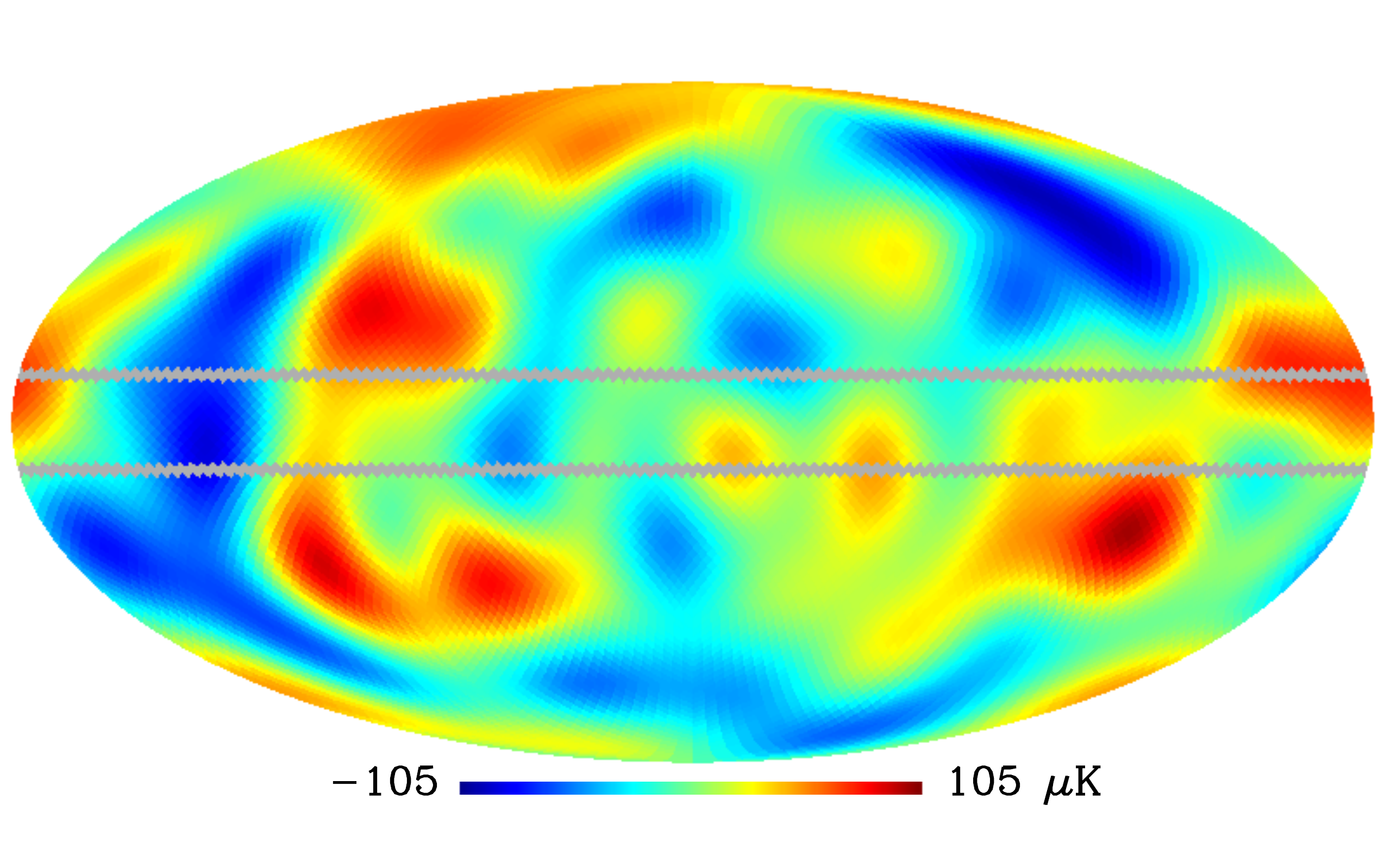}} \hfill
  \subfigure[final sparsity step]{\includegraphics[width=0.49\textwidth]{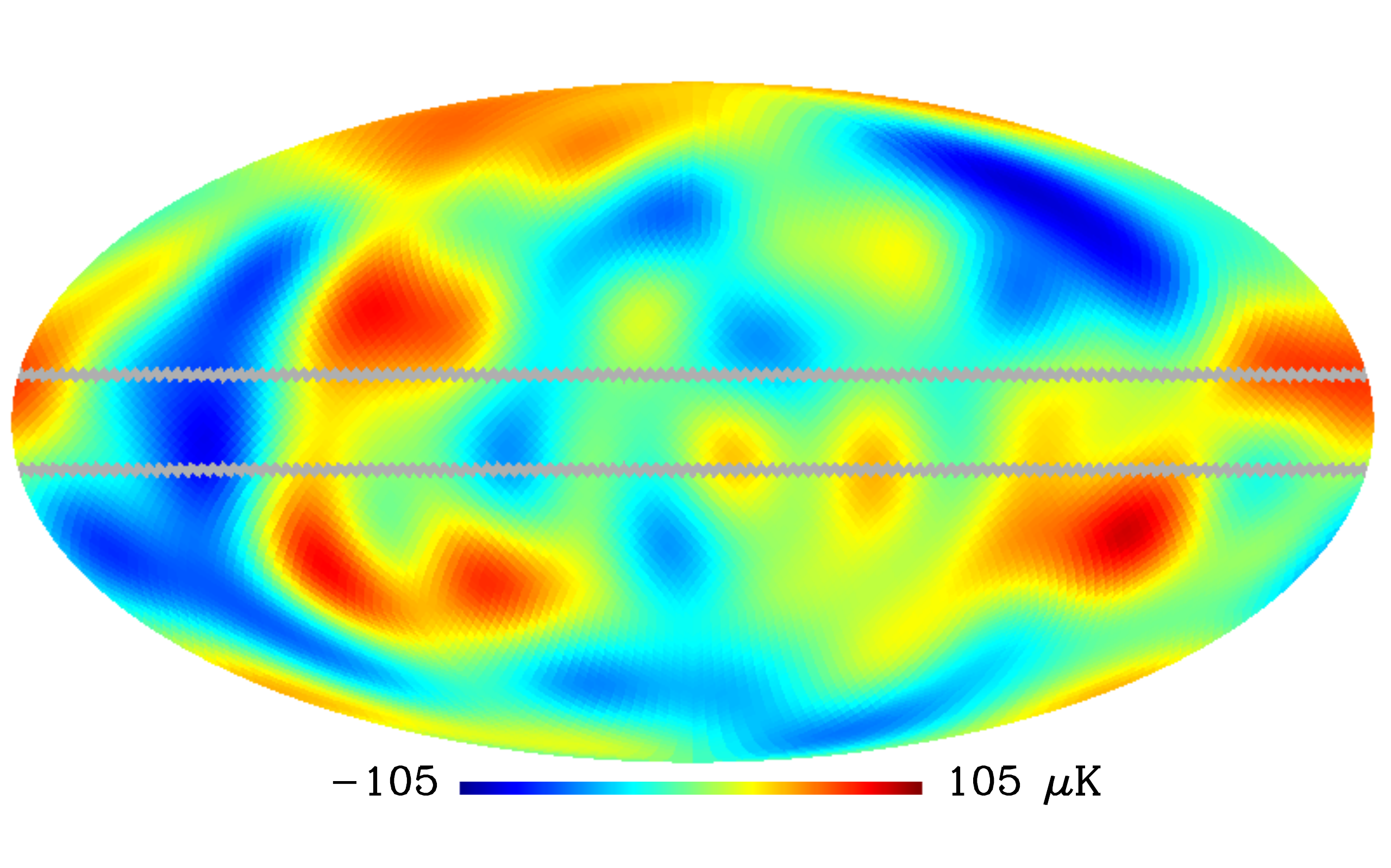}}
  \caption{Illustration of the sparse inpainting procedure. A Gaussian CMB realisation with $\lmax = 10$ (top left) is masked with a 20 degree sky cut (indicated in grey) and inpainted using the Douglas-Rachford algorithm. The evolution of the one-norm of the $\alm$s, $\| \vec{a} \|_1$, (light blue, solid) and the squared reconstruction error, $\| \vec{d} - \mtrx{Y} \vec{a}\|_2^2$ (dark-blue, dashed), are plotted in the top right, clearly showing the interplay between the two steps. The inpainted map is shown after the final data constraint step (bottom left) and the final sparsity step (bottom right).}
\label{fig:inpainting_demo}
\end{figure}

The interplay between the two steps in the Douglas-Rachford algorithm is 
illustrated in Fig.~\ref{fig:inpainting_demo}. In this demonstration, a 
Gaussian CMB map with $\lmax = 10$ (Panel (a)) is masked with a 20 degree sky cut
and inpainted using the joint sparse inpainting algorithm. In Panel (b) we show
the one-norm of the $\alm$s, $\| \vec{a} \|_1$, and the square of the reconstruction 
error, $\| \vec{d} - \mtrx{Y} \vec{a}\|_2^2$, each normalized by their maximum 
value, at each step of each iteration. It is clearly seen that the algorithm 
alternates between promoting data fidelity (at each ``half'' step) and sparsity 
(at each ``full'' step). Panels (c) and (d) show the inpainted map after the final 
data-constraint step and the final sparsity step: the fidelity of the reconstruction 
is higher in Panel (c), and the $\alm$s have lower magnitudes in Panel (d).

\section{Independent  Coefficients, Gaussianity and Isotropy}
\label{sec:gaussianity}

In this section we shall point out that independent, non-Gaussian 
spherical harmonic coefficients necessarily correspond to an 
anisotropic random field. It is proved by refs.~\cite{baldi:2006,baldi:2007} (see also ref.~\cite{marinucci:2011}, Ch.~5, and ref.~\cite{baldi:2013}) that, for an isotropic
random field, the $a_{\ell m}$ coefficients can be independent if
and only if they are Gaussian. While we do not reproduce the full
proof of this statement here, it is simple to provide a heuristic
explanation. It is indeed well
known that, under isotropy,%
\begin{equation*}
\vec{a}_{\ell }\overset{d}{=}D_{\ell }(g) \, \vec{a}_{\ell }\text{ ,}
\end{equation*}%
where $g\in \textrm{SO}(3)$ is any rotation in $\mathbb{R}^{3},$ $D_{\ell
}(g)$ represents Wigner's $D$-matrices (see, e.g.,
refs.~\cite{varshalovich:1989,marinucci:2011}), $\vec{a}_{\ell }$ represents the vector of
spherical harmonic coefficients at any multipole $\ell$, and
$\overset{d}{=}$ denotes equality in distribution, i.e., the left-
and right-hand sides have the same probability law. This identity
must hold for any change of coordinates under isotropy, but it is
easily seen to be violated by non-Gaussian independent priors.  For
instance, if we take $\vec{a}_{\ell}$ to have independent exponential
entries, then the components of $D_{\ell }(g)\vec{a}_{\ell }$ will be
represented by linear combinations of these exponentials; the
latter are neither independent nor exponentially distributed, for
general choices of the rotation $g$.

To give a more concrete example, let us assume we have generated a
random field by sampling independent Laplacian-distributed
$a_{\ell m}$s, which we take to have unit variance for notational simplicity
(see Appendix~\ref{sec:appendix}). Consider the multipole component%
\begin{equation*}
T_{\ell }(x)=\sum_{m}a_{\ell m}Y_{\ell m}(x)\text{ ,}
\end{equation*}%
with $x\in \textrm{S}^{2}$.  Because the spherical harmonics $Y_{\ell m}$ are identically zero
for $m\neq 0$
at the North Pole $N=(0,0,1)$, we have immediately%
\begin{equation*}
T_{\ell }(N)=a_{\ell 0}\sqrt{\frac{2\ell +1}{4\pi }}\text{ ,}
\end{equation*}%
meaning that at the North Pole $T_{\ell }(x)$ has itself a Laplacian
distribution. However, for different directions in the sky the law of
$%
T_{\ell }(x)$ will be given by linear combinations of the $a_{\ell
  m}$s weighted by spherical harmonics, and as such it will be much
closer to Gaussianity due to a central-limit theorem (CLT)-like
argument. For instance, the expected value of $T_{\ell }^{4}(x)$
(i.e., the trispectrum) will not be invariant to rotations, and isotropy
will consequently be violated. It is a matter of simple algebra to
show that
\begin{equation*}
\mathbb{E} \bigl( T_{\ell }^{2}(x) \bigr)=\Bigl\langle T_{\ell
}^{2}(x)\Bigr\rangle =\sum_{m}\left\vert Y_{\ell m}(x)\right\vert
^{2}\frac{1}{\sqrt{2}}\int_{-\infty }^{\infty
}u^{2}e^{-\sqrt{2}|u|}{\rm d}u=\frac{2\ell +1}{4\pi }\text{ ,}
\end{equation*}%
which shows that the variance of these multipole components is
indeed constant across the sky, so isotropy is not violated at the
second order level. However, for the fourth-moment we obtain%
\begin{equation*}
\mathbb{E} \bigl( T_{\ell }^{4}(N) \bigr) = \left\{ \sqrt{\frac{2\ell +1}{4\pi
}}\right\}^{4}
\expectation \bigl( a_{\ell 0}^{4} \bigr)
=\left\{ \sqrt{\frac{2\ell +1}{4\pi }}%
\right\} ^{4}\frac{1}{\sqrt{2}}\int_{-\infty}^{\infty
}u^{4}e^{-\sqrt{2}|u|}{\rm d}u =6\left\{ \sqrt{\frac{2\ell
+1}{4\pi }}\right\} ^{4}\text{ ,}
\end{equation*}%
so that, normalizing the variance to unity,
\begin{equation}
\label{eqn:eqn1}
\frac{\expectation \bigl( T_{\ell }^{4}(N) \bigr)}
{\Bigl[\expectation \bigl( T_{\ell }^{2}(N) \bigr)\Bigr] ^{2}}=6\text{ .}
\end{equation}%
More generally, for an arbitrary direction $x\in S^{2}$ we have
\begin{align*}
\mathbb{E(}T_{\ell }^{4}(x)) &=\mathbb{E}\Biggl\{
\sum_{m}a_{\ell m}Y_{\ell m}(x)\Biggr\} ^{4} \\
&=\sum_{m}\mathbb{E}\bigl(a_{\ell m}^{4}\bigr)\left\vert Y_{\ell
m}(x)\right\vert ^{4}+3\sum_{m,m^{\prime }}
\mathbb{E}\bigl(a_{\ell m}^{2}\bigr)\left\vert Y_{\ell m}(x)\right\vert
^{2} \mathbb{E}\bigl(a_{\ell m^{\prime }}^{2}\bigr)\left\vert
Y_{\ell m^{\prime }}(x)\right\vert ^{2} \\
&=6\sum_{m}\left\vert Y_{\ell m}(x)\right\vert ^{4}+3\sum_{m\neq
m^{\prime }} \left\vert Y_{\ell m}(x)\right\vert ^{2} \left\vert
Y_{\ell m^{\prime }}(x)\right\vert ^{2} \text{ }.
\end{align*}%
This quantity is clearly not constant for different values of $x$,
whence the anisotropic behaviour of fourth-order moments
(trispectra) is evident. In the next section, we provide
more detailed analysis on this same issue, and demonstrate that
features present in the trispectrum of pure-Laplacian random fields
can be inherited by sparsely inpainted Gaussian CMB maps.


\section{Analytic Computations and Simulations}
\label{sec:results}

\begin{figure}
\centering
  \subfigure[$\ell=2$]{\includegraphics[width=0.49\textwidth]{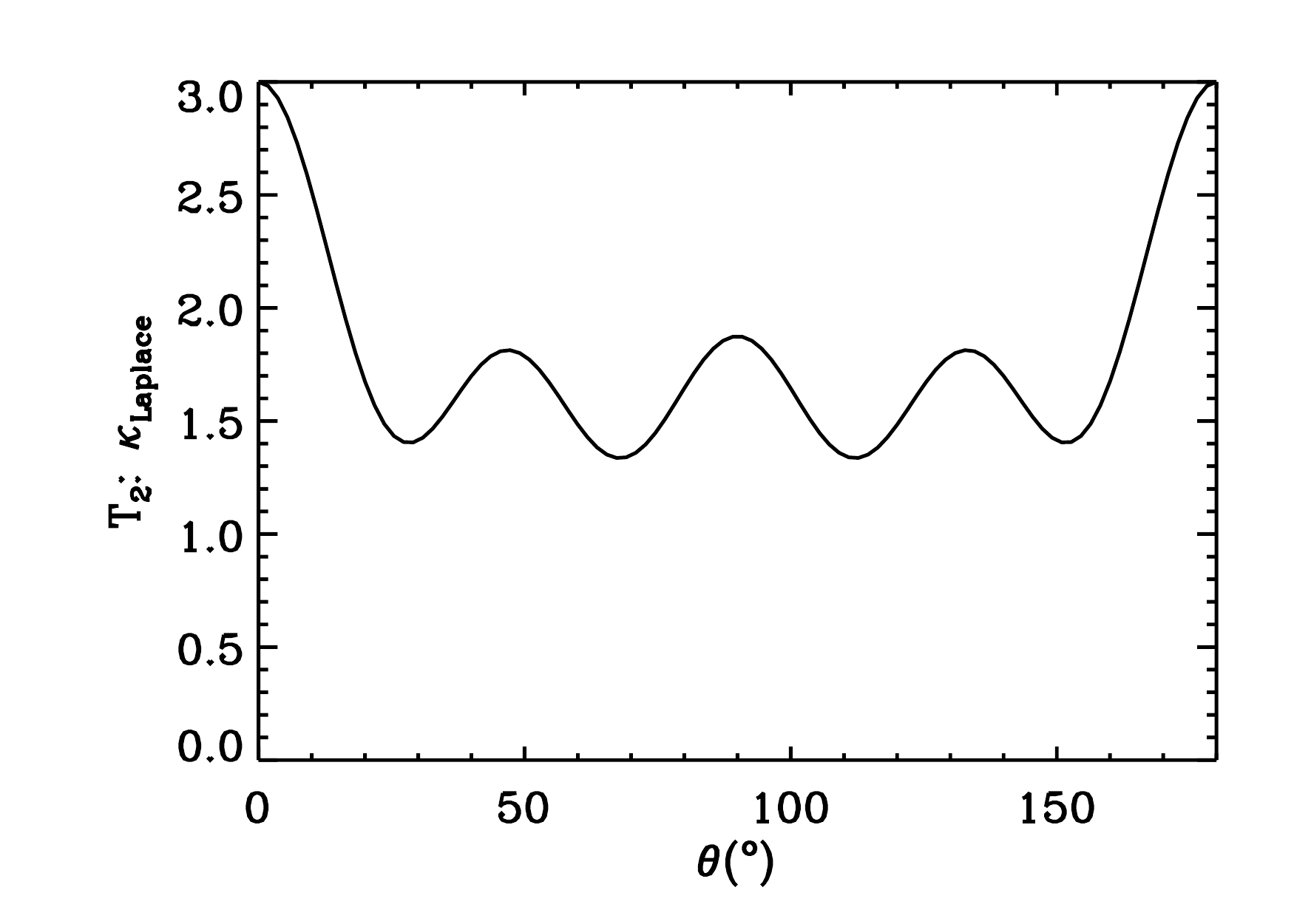}} \hfill
  \subfigure[$\ell=3$]{\includegraphics[width=0.49\textwidth]{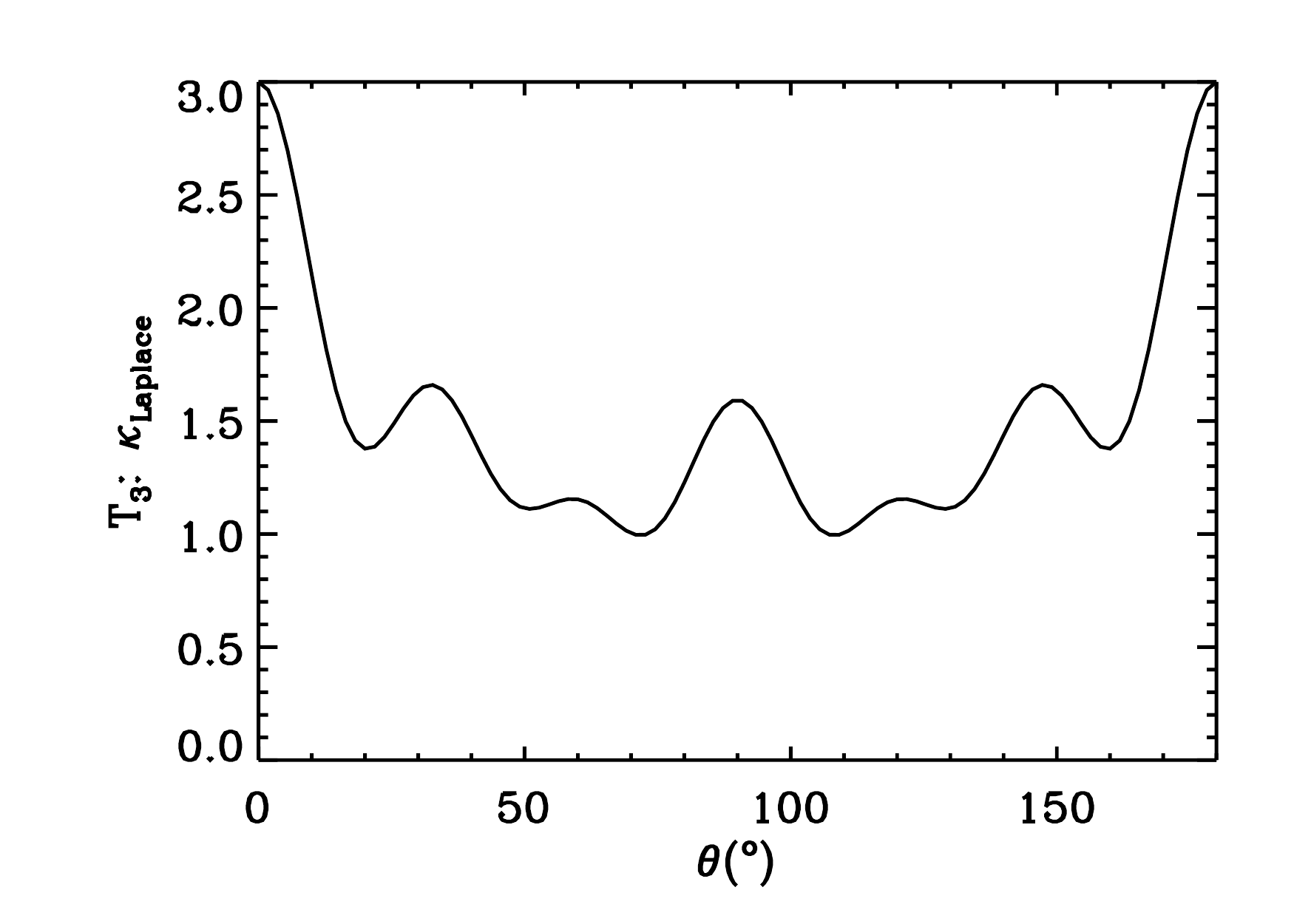}}
  \caption{Analytic Laplacian trispectrum as derived in
      Appendix~\ref{sec:appendix} for the separate inpainting optimization scheme. We plot the excess
      kurtosis evaluated on a grid of colatitude $\theta$, which is
      related to eqn.~(\ref{eqn:result1}) ($\ell=2$) and
      eqn.~(\ref{eqn:result2}) ($\ell=3$) via $\kappa =
      \expectation \bigl(T_\ell^4(\theta,0) \bigl) /
      \bigl[\expectation \bigl( T_{\ell }^{2}(\theta,0) \bigr)\bigr]
      ^{2} - 3$.}
\label{fig:analytic_trispectrum}
\end{figure}

In Appendix~\ref{sec:appendix} we provide analytic expressions
for the trispectrum of the quadrupole and the octupole in the case
of independent, Laplacian-distributed random spherical harmonic
coefficients $a_{\ell m}$. These expressions can be plotted on the
meridian with longitude $\varphi=0$, showing a clear modulation
over the sky with maxima at the North and South Poles and 
oscillatory behaviour between: see Fig.~\ref{fig:analytic_trispectrum}. 
The quantity plotted here is the excess kurtosis evaluated on a grid of
colatitude $\theta$, which is related to eqn.~(\ref{eqn:result1})
($\ell=2$) and eqn.~(\ref{eqn:result2}) ($\ell=3$) given in
Appendix~\ref{sec:appendix} via $\kappa = \expectation
\bigl(T_\ell^4(\theta,0) \bigl) / \bigl[\expectation \bigl(
T_{\ell }^{2}(\theta,0) \bigr)\bigr] ^{2} - 3$.

As shown by our simulations below, this pattern can remain in
Gaussian inpainted maps, thus introducing a small but
non-negligible deviation from isotropy. To demonstrate this effect, we
have generated $100\,000$ Gaussian isotropic maps at low resolution
($\lmax=10$,\footnote{The results of ref.~\cite{starck:2013b} were
computed using a higher band-limit of 50. Due to the large number of simulations 
required to demonstrate our results, we use $\lmax = 10$; we have 
verified that our results do not change when $\lmax$ is set to 50.} 
$\nside = 32$). We mask a 60-degree-wide
azimuthally-symmetric strip, centered on the equator,
and inpaint the masked maps using the $\ell_{1}$-minimization
procedure described in Sect.~\ref{sec:inpainting}, employing both
joint and separate sparsity priors. To ensure that a 
satisfactory solution to the constrained optimization problem 
(eqn.~\ref{eqn:optimization_problem_2}) is found for each map,
we augment this procedure with a convergence check, as opposed to running 
for a set number of iterations. The algorithm is taken to have converged 
when the energy of the update to the inpainted $a_{\ell m}$s at the $n^{\rm th}$ 
iteration, $\Delta \vec{a}^n$, is less than 0.1\% of the energy of the 
inpainted $a_{\ell m}$s themselves, i.e., when $\| \Delta \vec{a}^n \|_2 /
\| \vec{a}^n \|_2  \le 0.001$.
If the convergence criterion is not satisfied within 150 iterations, we 
replace the map with a new random draw and repeat the masking 
and inpainting procedure. To gain insight into the performance of the 
inpainting procedure, we also calculate two other estimates of the 
full-sky $a_{\ell m}$s for each of the $100\,000$ input maps, namely the 
cut-sky $a_{\ell m}$s and the Wiener-filtered $a_{\ell m}$s. Finally, we 
evaluate the Monte Carlo averages for the trispectrum of the
quadrupole and octupole for all four estimators in the form of the excess 
kurtosis along the $\varphi = 0$ meridian: $\kappa =  \expectation 
\bigl(T_\ell^4(\theta,0) \bigl) / \bigl[\expectation \bigl( T_{\ell }^{2}(\theta,0) 
\bigr)\bigr]^{2} - 3$.

\begin{figure}
\centering
  \subfigure[$\ell=2$]{\includegraphics[width=0.49\textwidth]{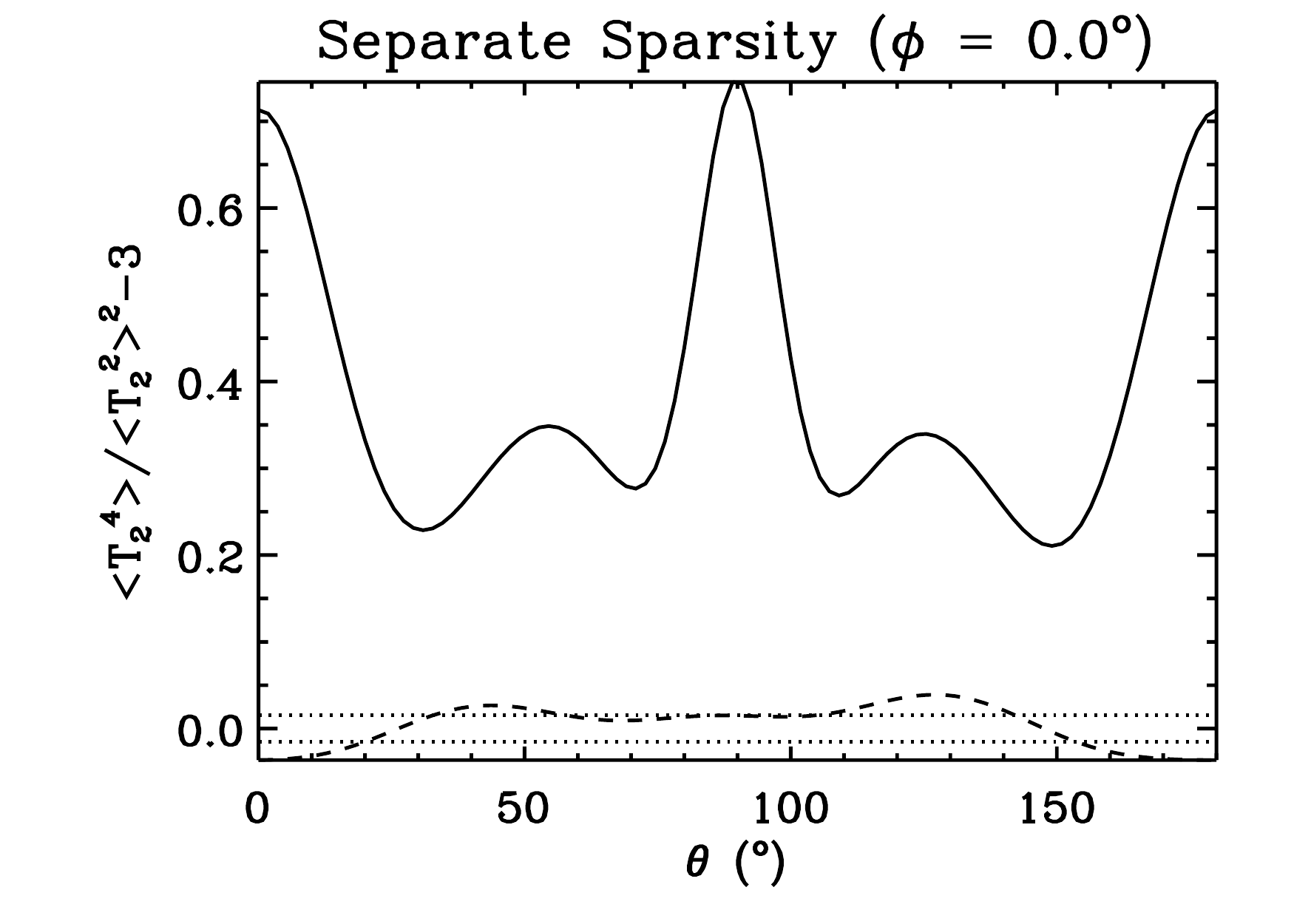}} \hfill
  \subfigure[ $\ell=3$]{\includegraphics[width=0.49\textwidth]{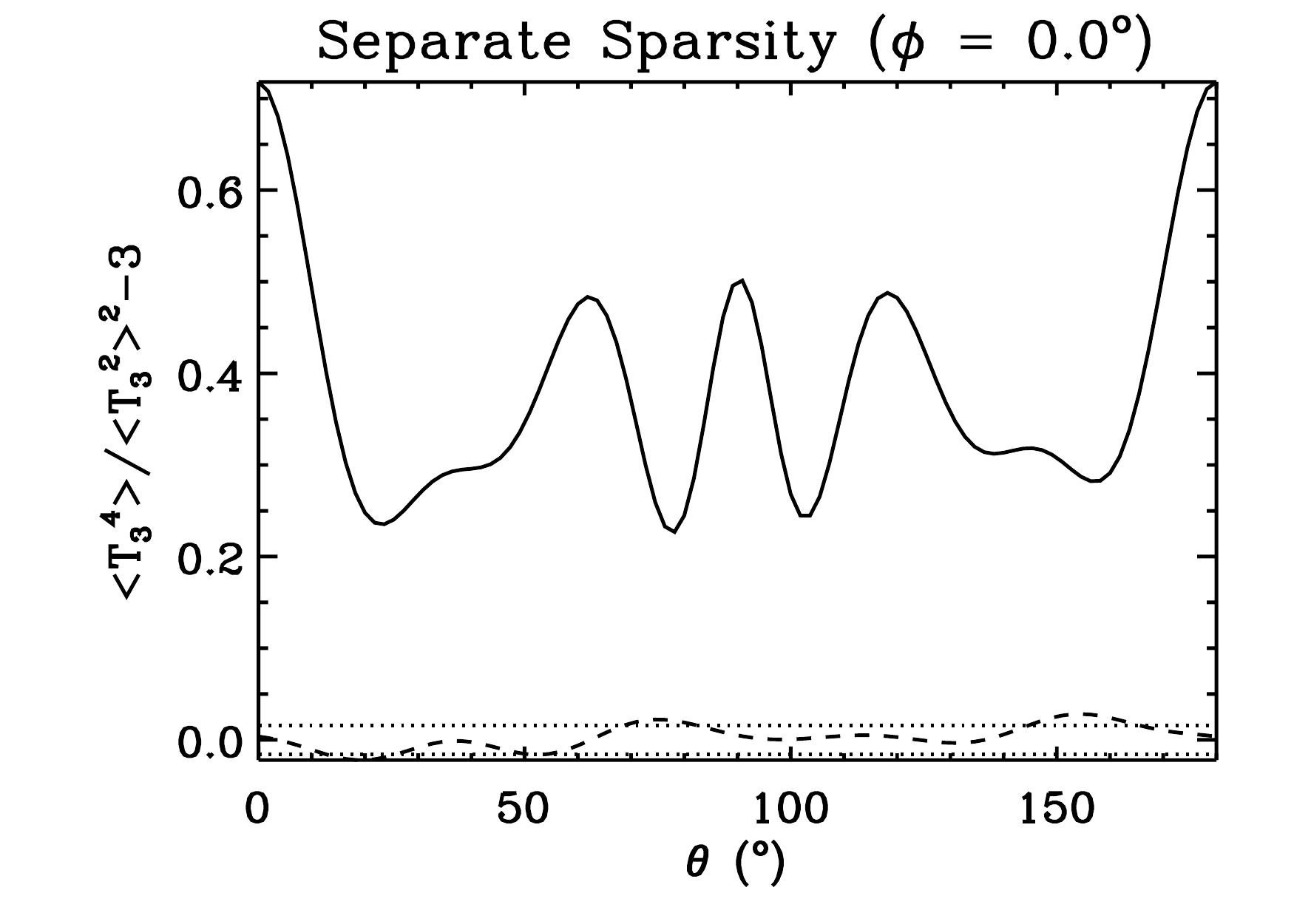}}
  \caption{Simulated inpainted trispectrum, assuming {\em separate} 
    sparsity.  The trispectrum of the ensemble of inpainted maps is 
    shown as a function of colatitude $\theta$ by the solid black line; 
    the trispectrum of the input Gaussian maps is shown by the 
    dashed line. The dotted lines show the one-sigma Monte Carlo 
    sampling error on the trispectrum of the $100\,000$ Gaussian 
    realizations, which are manifestly consistent with Gaussianity 
    and isotropy. The trispectrum of the inpainted maps is clearly significantly
    different from zero everywhere, and is similar to the Laplacian
    trispectrum analytically derived in Appendix \ref{sec:appendix}
    and plotted in Fig.~\ref{fig:analytic_trispectrum}.  The magnitude
    of the anisotropic signal for a given $\ell$-mode is around
    25\% of the Gaussian expected value. Results are obtained using 
    input maps with $\lmax=10$, and cutting a 60 degree region, i.e., one
    half of the sky. }
\label{fig:simulated_trispectrum_sep_60}
\end{figure}

\begin{figure}
\centering
  \subfigure[$\ell=2$]{\includegraphics[width=0.49\textwidth]{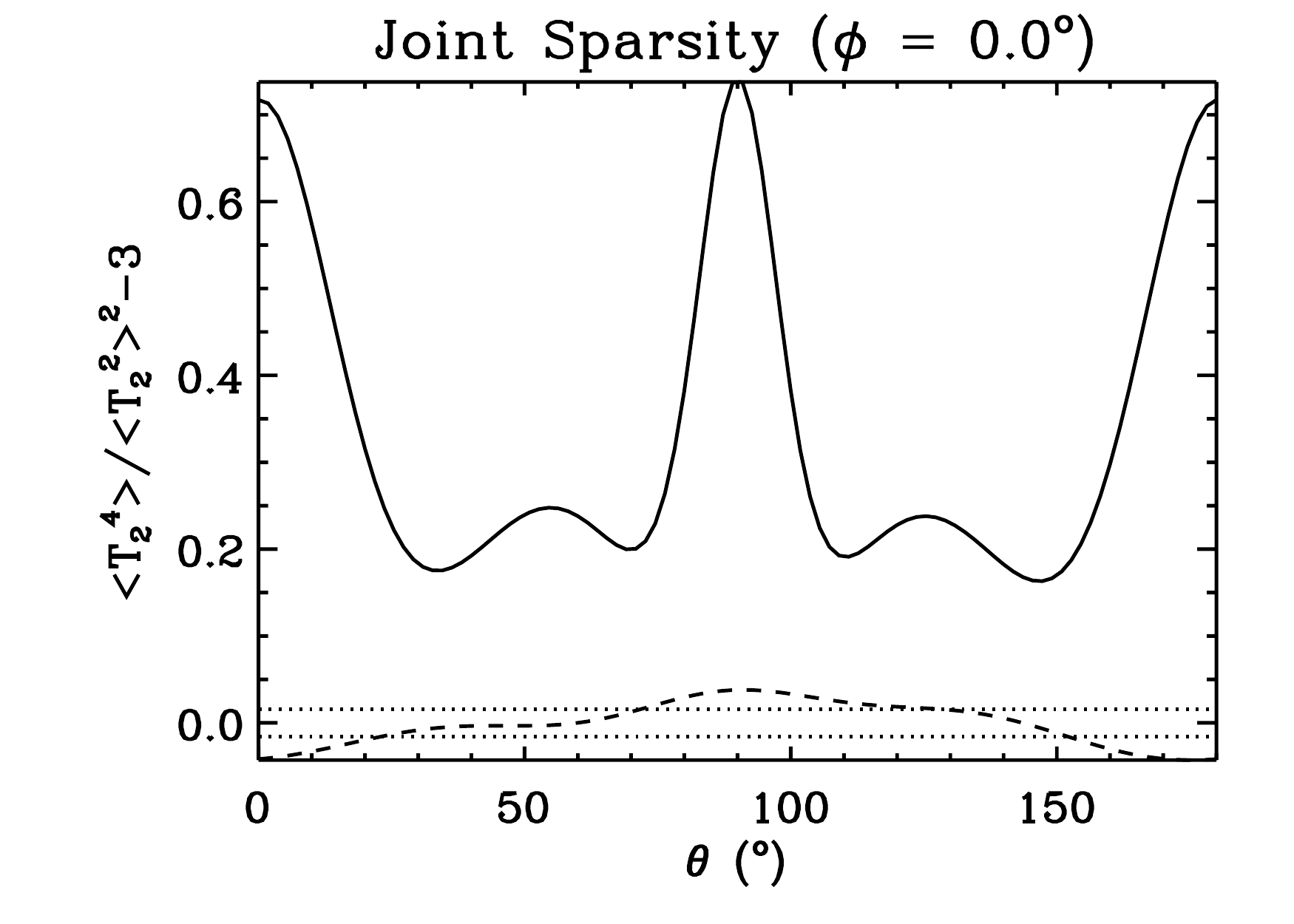}} \hfill
  \subfigure[ $\ell=3$]{\includegraphics[width=0.49\textwidth]{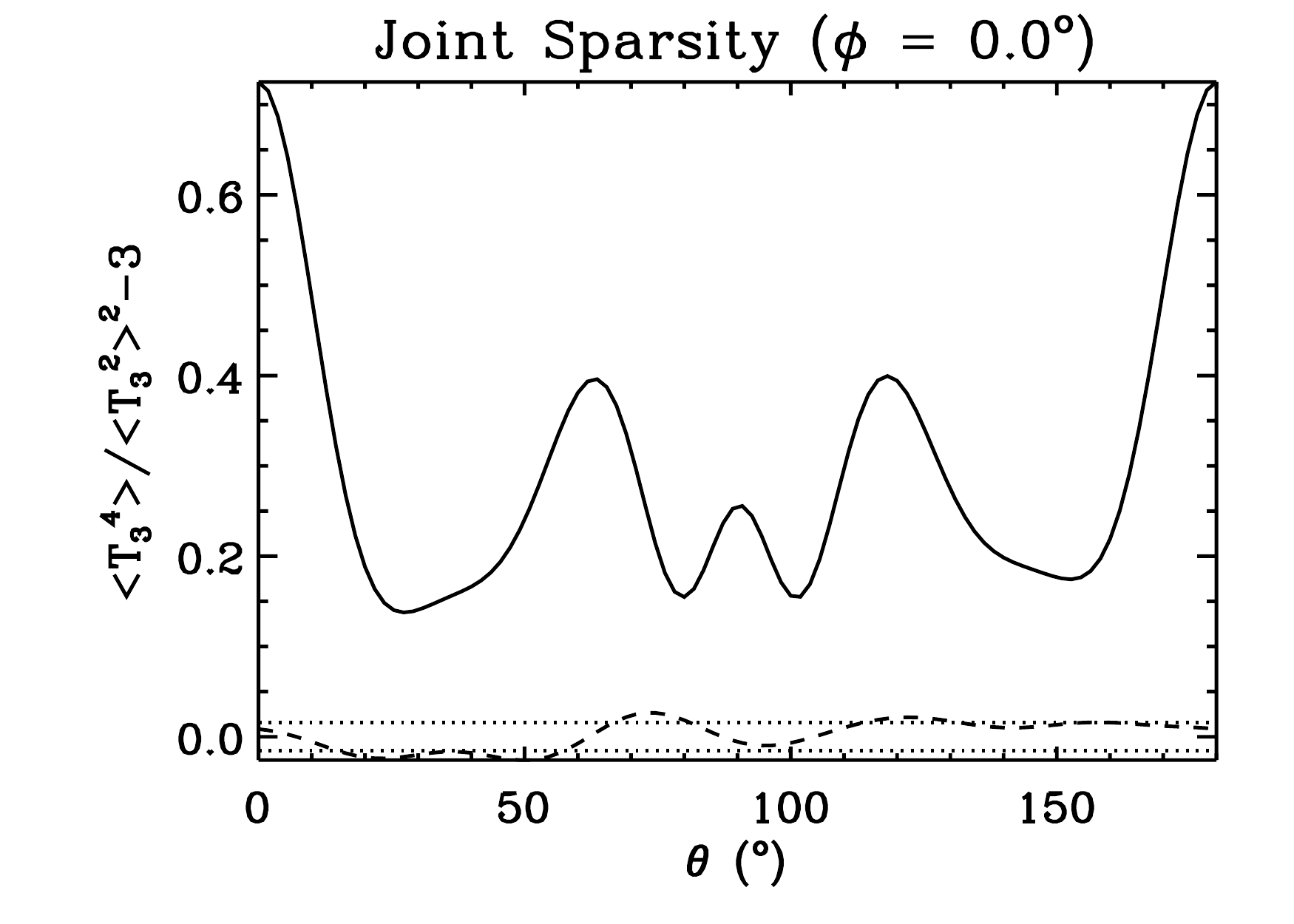}}
  \caption{Simulated inpainted trispectrum, assuming {\em joint} 
    sparsity, using the same line conventions as Fig.
    \ref{fig:simulated_trispectrum_sep_60}. The trispectrum of the inpainted maps is, again, 
    clearly significantly different from zero everywhere; the precise
    shape of the trispectrum of the inpainted maps is slightly different 
    to the separate-sparsity case, but the magnitude of the anisotropic 
    signal for a given $\ell$-mode remains $\sim25\%$ of the Gaussian 
    expected value. As before, results are obtained using input maps with
    $\ell_{\mathrm{max}}=10$, and cutting a $60^\circ$ region, i.e., one
    half of the sky.   }
\label{fig:simulated_trispectrum_joint_60}
\end{figure}

\begin{figure}
\centering
  \subfigure[$\ell=2$]{\includegraphics[width=0.49\textwidth]{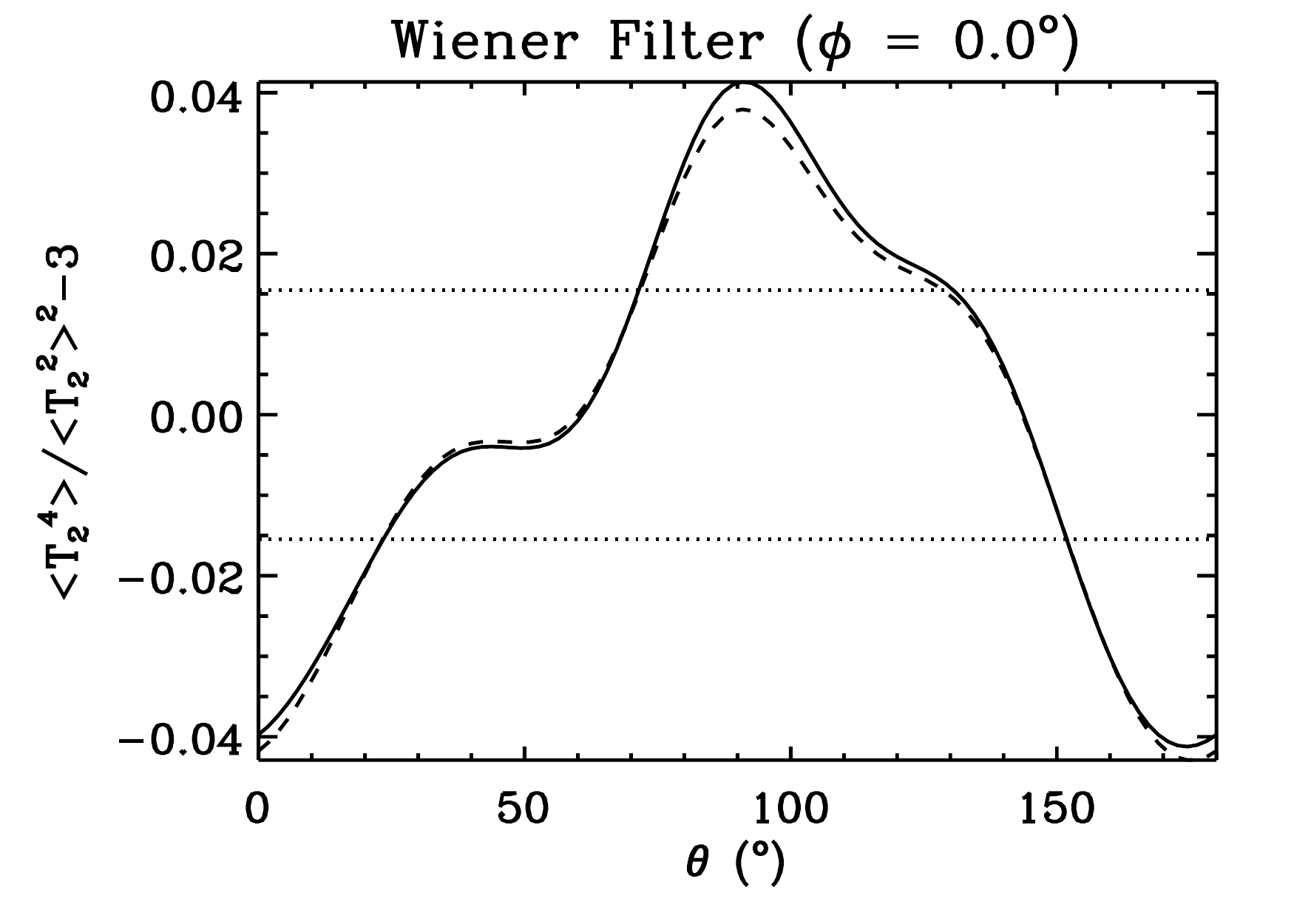}} \hfill
  \subfigure[ $\ell=3$]{\includegraphics[width=0.49\textwidth]{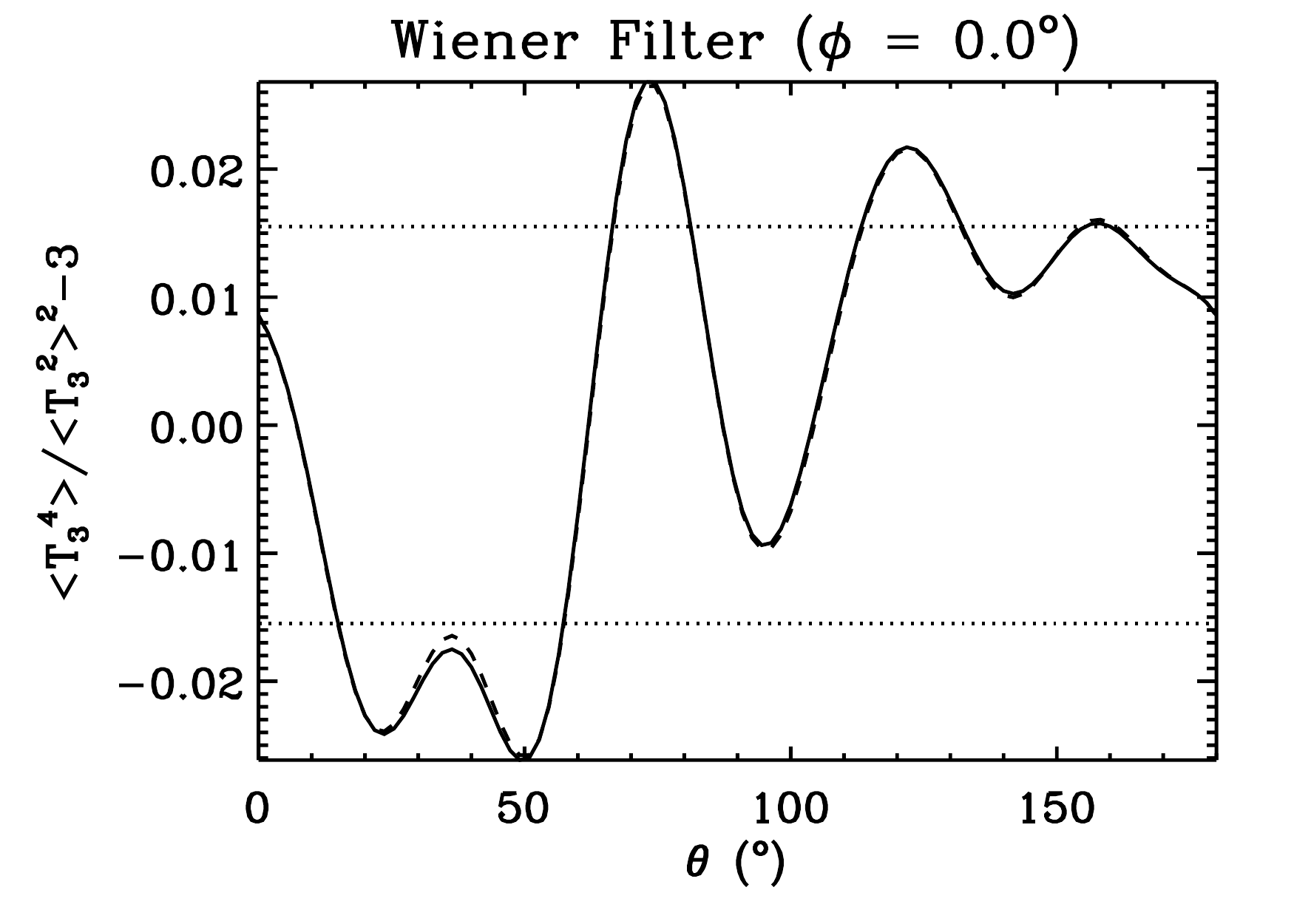}}
  \subfigure[$\ell=2$]{\includegraphics[width=0.49\textwidth]{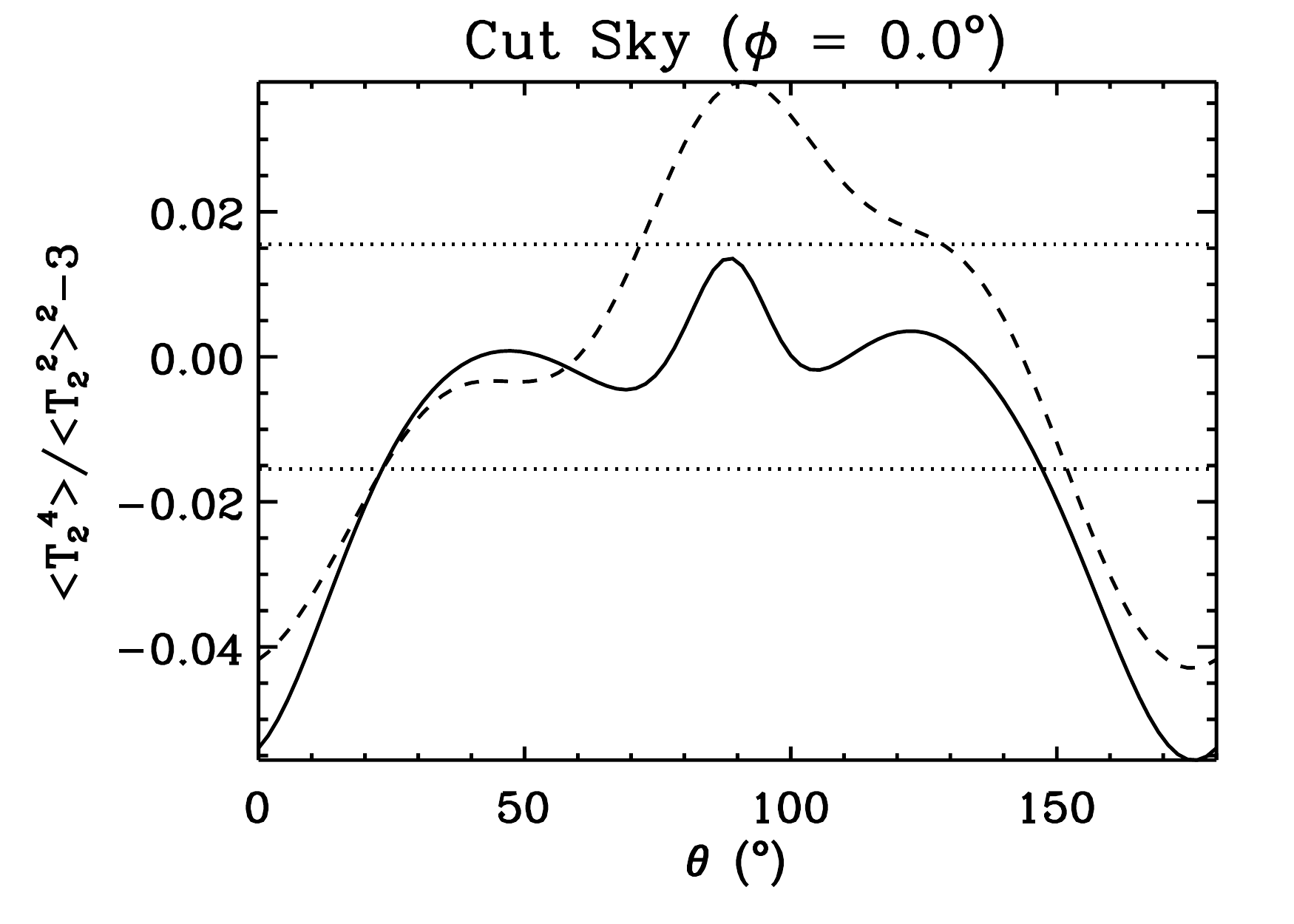}} \hfill
  \subfigure[ $\ell=3$]{\includegraphics[width=0.49\textwidth]{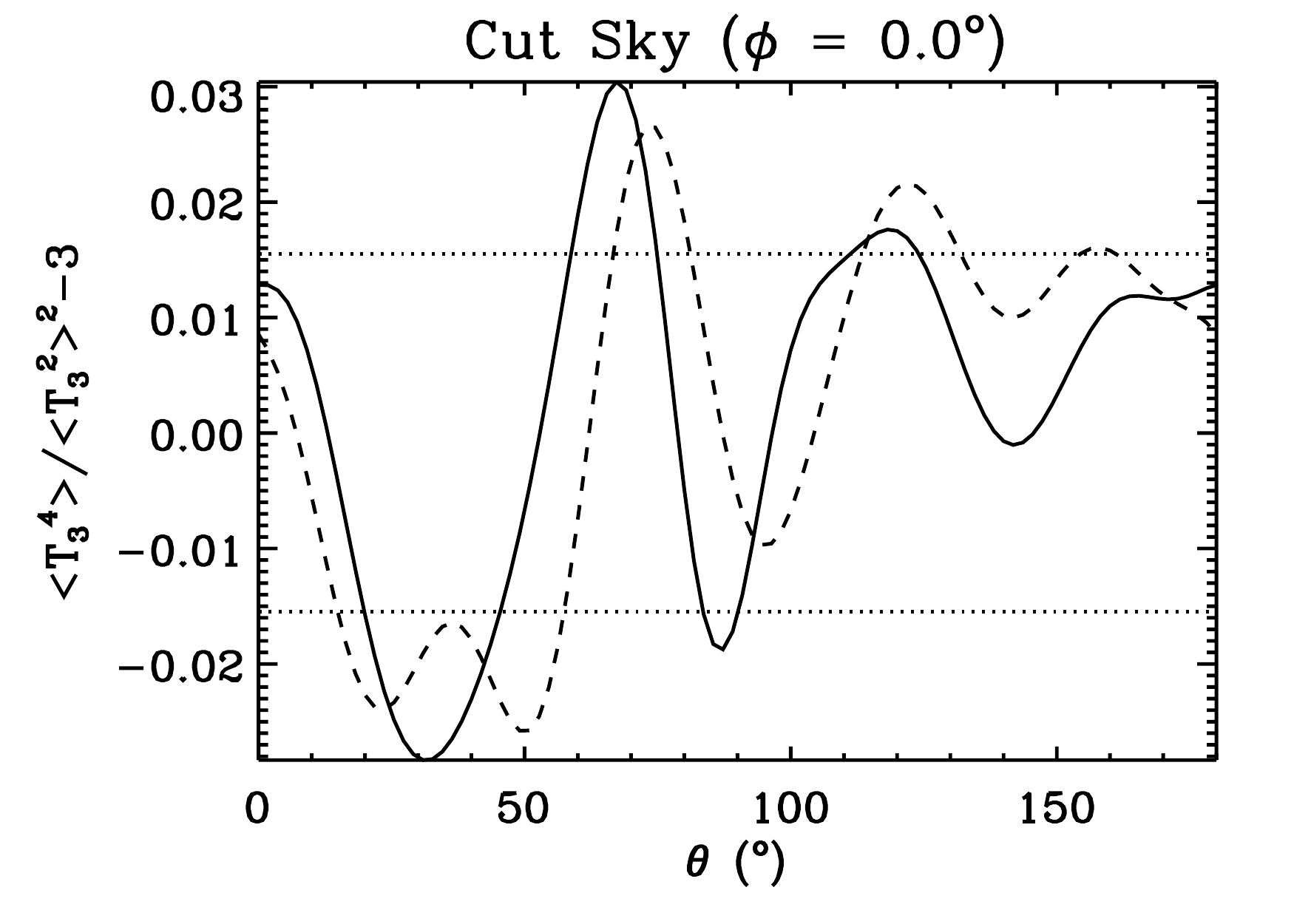}}
  \caption{Trispectra of Wiener-filtered (top) and cut-sky (bottom) maps, 
    using the same line conventions as Fig.~\ref{fig:simulated_trispectrum_sep_60}. 
    Neither the Wiener-filtered nor cut-sky 
    maps show any significant deviation from isotropy.}
\label{fig:simulated_trispectrum_wf_cs_60}
\end{figure}

Our primary results are shown in Figs.~\ref{fig:simulated_trispectrum_sep_60},
\ref{fig:simulated_trispectrum_joint_60} and~\ref{fig:simulated_trispectrum_wf_cs_60}. 
The main finding is summarized by the solid black line in each plot: this is the trispectrum of
the ensemble of inpainted maps (for $\ell = 2$ or $\ell = 3$), evaluated on a grid of
colatitude $\theta$. Also plotted 
are the trispectrum of the input Gaussian maps (dashed), and the one-sigma 
Monte Carlo sampling error on the trispectrum of the Gaussian maps, for 
$100\,000$ realizations. The trispectrum of the input maps is clearly consistent with 
Gaussianity and isotropy. On the other hand, it is quite evident from
Figs.~\ref{fig:simulated_trispectrum_sep_60} and~\ref{fig:simulated_trispectrum_joint_60}
that the trispectra of the sparse-inpainted maps are clearly significantly different from zero
everywhere, and exhibit similar qualitative features as those
illustrated in our analytic computations (c.f. Fig.~\ref{fig:analytic_trispectrum}): namely a clear
anisotropic modulation over the meridian, with peaks at the North
Pole for both the quadrupole and octupole. Note that the maximum
anisotropy appears at the points furthest from the mask: hence the effect we are
describing cannot simply be confined within the inpainted portion
of the map.

This anisotropy is not present in the maps generated using the other $a_{\ell m}$ 
estimators. Fig.~\ref{fig:simulated_trispectrum_wf_cs_60} shows no significant 
non-Gaussianity nor anisotropy for either the cut-sky or Wiener-filtered maps:
 the fluctuations are fully within the Monte Carlo one-sigma confidence bounds,
represented by the dotted lines. As far as the cut-sky maps are
concerned, this may at first sight seem somewhat surprising, given
that the mask does introduce anisotropic behavior;  however, this
can be explained by noting that the anisotropy in the variance is
fully taken into account, because the kurtosis is itself
normalized by the Monte Carlo second moment at the corresponding
colatitude.

We recall that the results are obtained with a sky cut masking a 60 degree
strip, i.e. half of the sky; larger than those used in typical CMB analyses. We 
have therefore repeated the analysis using a smaller sky cut, removing only 
a 20 degree strip of sky, and have found a similar pattern in the trispectra 
of the inpainted maps, albeit at reduced amplitude. The
magnitude of the anisotropic signal for a given $\ell$-mode is
around 25\% of the expected Gaussian value for the 60 degree
sky cut, and 3--4\% for a
smaller sky cut of 20 degrees.\footnote{Note that it is also possible to 
terminate the inpainting procedure after the {\em second} step of the 
final iteration, eqn. (\ref{eqn:dr_prox_step}), rather than the {\em first step},
eqn. (\ref{eqn:dr_proj_step}). The magnitude of the anisotropic 
signal increases in this case to $\sim30\%$ for the 60 degree mask and
$\sim6\%$ for the 20 degree mask.} Nevertheless, the effect is highly
significant for the present simulations: the signal is several
times larger than the Monte Carlo variance. Note also that, while
the effect for a given multipole configuration is small, the
cardinality of different trispectra configurations at {\em Planck}
\citep{PlanckI:2013} resolution is very large. Thus, it is possible 
that statistics can be formulated that will spuriously detect broken 
statistical isotropy in inpainted maps due to the effect described in 
this work; for instance, an anisotropic feature may be mistakenly 
confused with an unexpected stochastic dependence. Further 
investigations are hence needed to assess the cumulative impact 
for the statistical analysis of a complete data set.

\begin{figure}
\centering
  \subfigure[joint sparsity]{\includegraphics[width=0.49\textwidth]{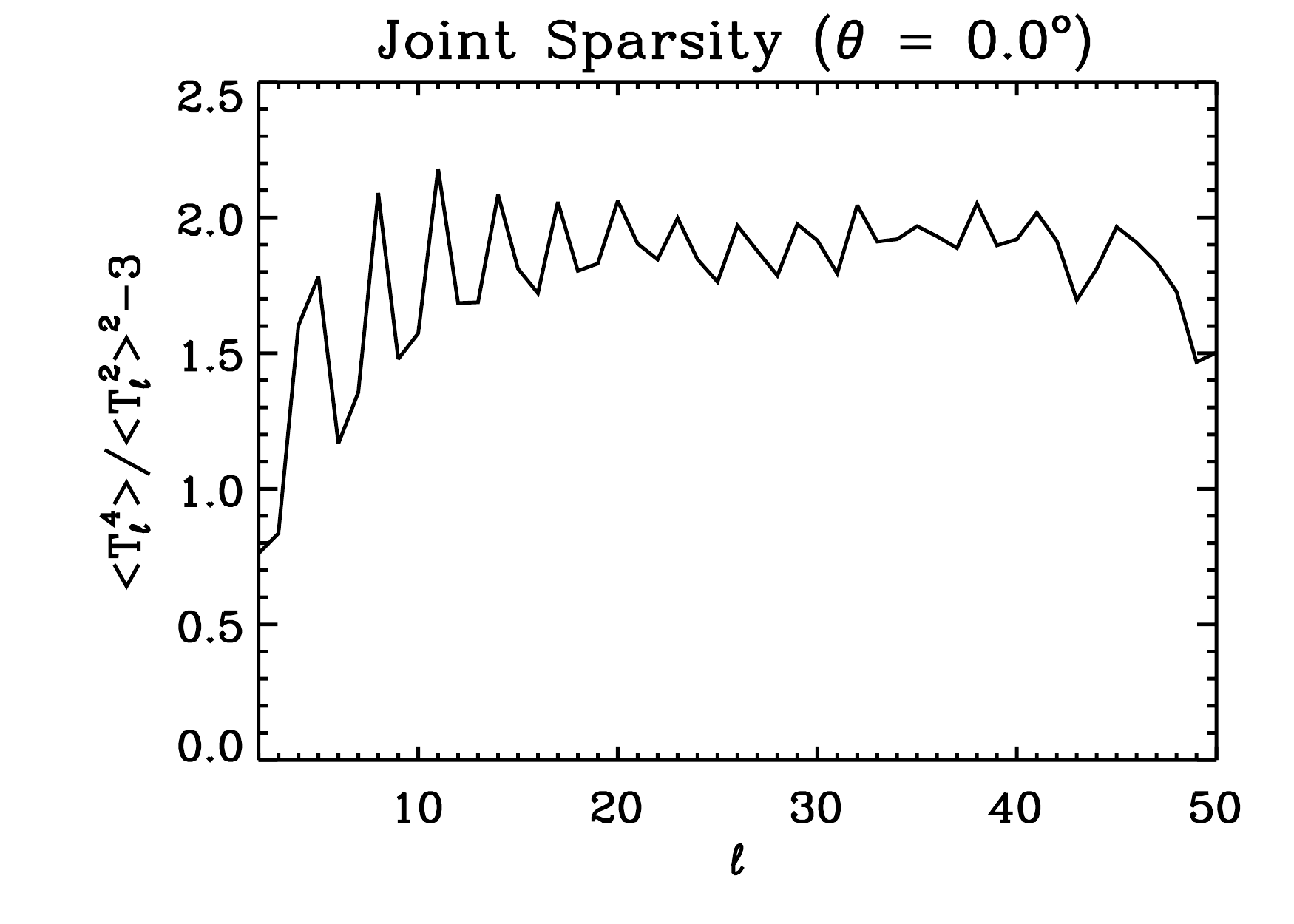}} \hfill
  \subfigure[separate sparsity]{\includegraphics[width=0.49\textwidth]{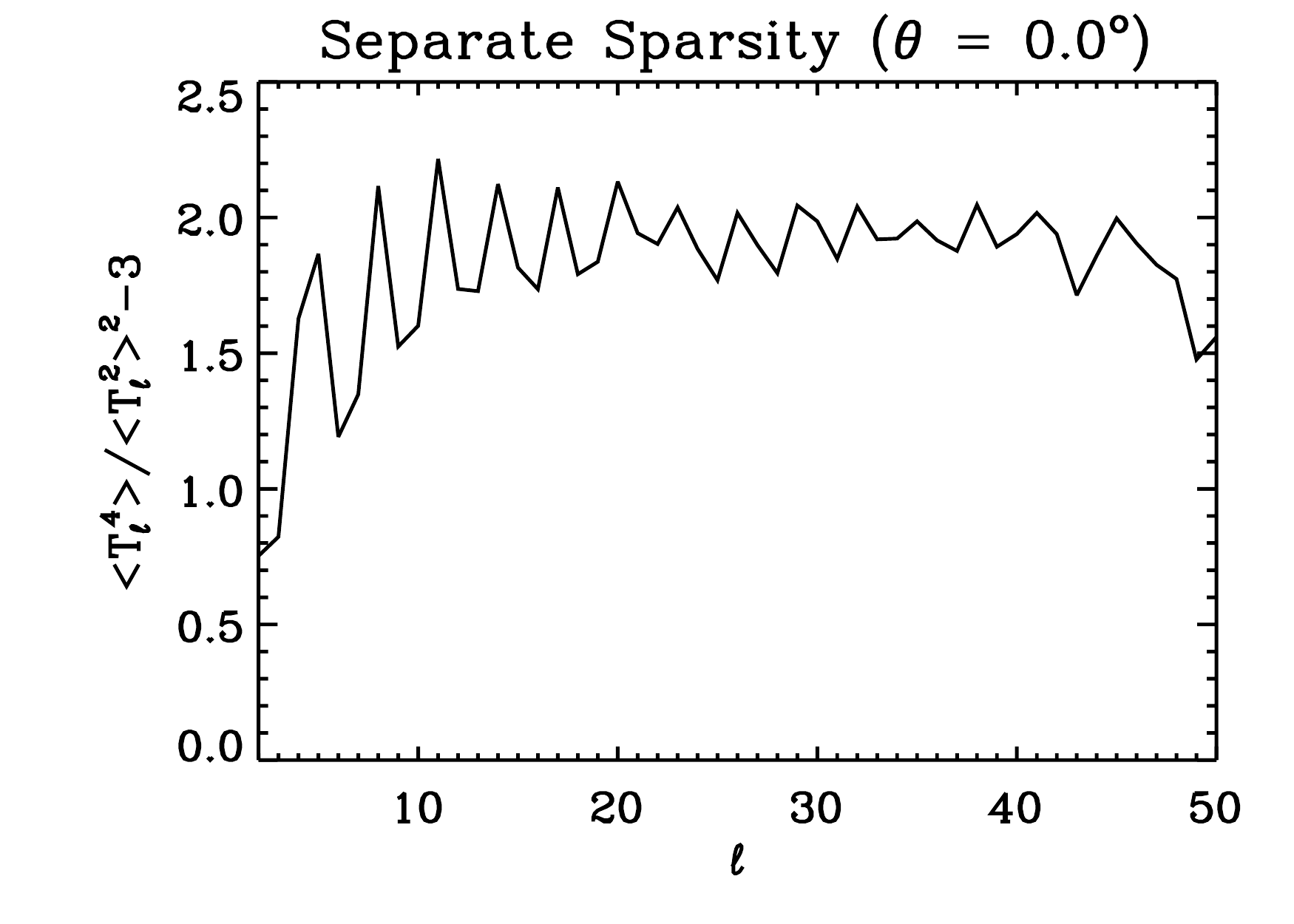}}
  \caption{Simulated inpainted trispectrum, evaluated at the North Pole
    only, for all multipoles up to 50. The results for both joint (left) and 
    separate (right) sparsity indicate that multipoles beyond the octupole
    are also made anisotropic by the inpainting procedure.}
\label{fig:simulated_north_pole_trispectrum_60_l_max_50}
\end{figure}

We also reiterate that the results are obtained using maps with a low 
band-limit ($\ell_{\mathrm{max}}=10$), due to the large number of 
simulations required to reduce the sampling error on the kurtosis. 
To provide some hints on the behaviour of trispectra at higher 
multipoles, we have repeated the simulation-inpainting procedure
using a higher band-limit of $\lmax=50$, calculating the kurtosis of 
$a_{\ell0}$. Recall that the spherical harmonics with $m \ne 0$ are 
zero at the North Pole, and thus the kurtosis of $T_\ell$ reduces to 
the kurtosis of $a_{\ell0}$ at this position. Fig.~\ref{fig:simulated_north_pole_trispectrum_60_l_max_50} shows the
kurtosis of $a_{\ell0}$ for $\ell=2-50$: it is clear from this figure that, 
for both forms of the sparsity prior, the effect does not vanish at 
higher $\ell$. Again, we leave for future work a complete investigation of 
these effects under more realistic experimental settings.

\section{Conclusions}
\label{sec:conclusions}

We have provided analytic computations and simulations suggesting
that sparse inpainting techniques using the spherical 
harmonic dictionary can alter the isotropy properties of
 Gaussian random fields on the sphere.  The effect for
a given multipole configuration is around 25\% of the Gaussian
expected value for a 50\% sky cut and remains significant for
smaller masks.  It must also be kept in mind that the cardinality
of different trispectra configurations is extremely high at {\em Planck}
resolution; further investigations are hence needed to assess the
cumulative impact for the statistical analysis of a full data set.
These issues are beyond the scope of the present paper and left as
an avenue for further research, as are investigations into the isotropy
properties of sparse inpainting using other dictionaries, such as wavelets.

As noted earlier, our Monte Carlo analysis does not take into
account some features of realistic CMB data analysis: in
particular, for brevity we focussed on low resolution maps, we did
not consider beam and noise effects, and we
investigated only partially the effect of a varying width for the
masked region. We believe, however, that these technicalities do
not affect the basic message of this paper, and we expect that our
main theoretical conclusions will remain substantially unaltered
under more realistic settings. As previously
mentioned, these issues and the relevance of our findings
for the analysis of data sets such as those from WMAP~\citep{wmap9:2012} and
\emph{Planck} are left as topics for future research.


\bibliographystyle{JHEP}
\bibliography{biblio}


\acknowledgments{
  We thank Jean-Luc Starck, Mike Hobson, Yves Wiaux, R\'{e}mi Gribonval
  and Mike Davies for useful and lively discussions. SMF is supported by STFC and a
  grant from the Foundational Questions Institute (FQXi) Fund, a
  donor-advised fund of the Silicon Valley Community Foundation on the
  basis of proposal FQXiRFP3-1015 to the Foundational Questions
  Institute. Research by DM and VC is supported by the European Research Council under the European
  Community Seventh Framework Programme (FP7/2007-2013) ERC grant
  agreement no.~277742 \emph{Pascal}. JDM is supported in part by a Newton International Fellowship
  from the Royal Society and the British Academy. HVP is supported by
  STFC, the Leverhulme Trust, and the European Research Council under
  the European Community Seventh Framework Programme (FP7/2007-2013) /
  ERC grant agreement no.~306478 \emph{CosmicDawn}. BDW is supported
  by the ANR Chaires d'Excellence programme and NSF grants AST
  07-08849 and AST 09-08902 during this work. This work was also
  supported by ASI/INAF Agreement I/072/09/0 for the {\em Planck} LFI
  Activity of Phase E2, French state funds managed by the ANR within
  the Investissements d'Avenir programme under reference
  ANR-11-IDEX-0004--02, and by the National Science Foundation under
  Grant No.~PHY11-25915.  The authors acknowledge the hospitality of the 
  Big Bang, Big Data, Big Computers workshop at APC, Paris. Some of the 
  results in this paper were derived using the HEALPix package \citep{gorski:2005}.}


\appendix
\section{The Laplacian trispectrum for the quadrupole and the
octupole}
\label{sec:appendix}

In this appendix, we provide some analytic computations to
illustrate the anisotropic features which can be introduced by
sampling independent, non-Gaussian $a_{\ell m}$s. It should be
noted that we use the prior corresponding to the separate sparsity
implementation eqn.~(\ref{eqn:l1norm1}), rather than the joint
one in eqn.~(\ref{eqn:l1norm2}). The latter would lead to a
different modulation pattern, but it would not alter the message
we want to deliver on the existence of anisotropic features. As we
have seen in the simulation section, the realized trispectra are
very much the same under both optimization schemes.

We make use of several analytic expressions for the spherical
harmonics $Y_{\ell m}$ and the associated Legendre functions
$P_{\ell m}$; we refer for instance to 
refs.~\cite{varshalovich:1989,marinucci:2011} for further discussion on these
functions.

More precisely, we shall focus on the quadrupole $\ell=2$ and
octupole $\ell=3,$ taking for simplicity $\varphi =0.$ For
symmetry, we use the real-valued spherical harmonic basis,
\begin{equation*}
\left\{ \mathcal{Y}_{\ell m},m=-\ell,...,\ell\right\} =\left\{ Y_{\ell 0},\text{ }\frac{
Y_{\ell m}+\overline{Y}_{\ell m}}{\sqrt{2}}\text{ , }\frac{Y_{\ell m}-\overline{Y}_{\ell m}}{
\sqrt{2}i},\text{ }m=1,...,\ell\right\} \text{ ,}
\end{equation*}
whence we have the equivalence
\begin{equation*}
T_{\ell }(\theta ,\varphi )=\sum_{m}a_{\ell m}Y_{\ell m}(\theta ,\varphi )=\sum_{m}\alpha _{\ell m}
\mathcal{Y}_{\ell m}\text{ .}
\end{equation*}
We take the coefficients $\left\{ \alpha _{\ell m}\right\} $ to be
real-valued and Laplacian-distributed with variance unity, so that
\begin{equation*}
\mathbb{E}[\alpha _{\ell m}]=\mathbb{E}[\alpha _{\ell m}^{3}]=0\text{ , }\mathbb{E}
[\alpha _{\ell m}^{2}]=1\text{ , }\mathbb{E}[\alpha _{\ell
m}^{4}]=6\text{ ; }
\end{equation*}
in terms of the complex-valued coefficients $\{a_{\ell m}\}$ this entails that
\begin{equation*}
\mathbb{E}[a_{\ell 0}]\text{ }=\mathbb{E}[a_{\ell 0}^{3}]=0\text{ },\text{ }\mathbb{E}
[a_{\ell 0}^{2}]=C_\ell=1\text{ },\text{ }\mathbb{E}[a_{\ell 0}^{4}]=6\text{
,}
\end{equation*}
while for $m\neq 0$,
\begin{equation*}
\mathbb{E}[a_{\ell m}]=0\text{ },\text{ }\mathbb{E}[{ \Re}{a_{\ell m}}]^{2}=
\mathbb{E}[ { \Im}{a_{\ell m}}]^{2}=\frac{C_\ell}{2}=\frac{1}{2}\text{ },\mathbb{E}[ { \Re}{
a_{\ell m}}]^{4}=\mathbb{E}[ { \Im}{a_{\ell
m}}]^{4}=\frac{3}{2}\text{ },
\end{equation*}
with $ {\Re}$ and $ {\Im}$ denoting as usual real and imaginary
parts. Some simple algebra, using the definition and parity properties of
spherical harmonics, yields

\begin{align*}
T(\theta ,\varphi )& =\sum_{\ell =0}^{\infty }\sum_{m=-\ell
}^{\ell}a_{\ell m}Y_{\ell m}(\theta ,\varphi )\\
&=\sum_{\ell =0}^{\infty
}\sum_{m=0}^{\ell}a_{\ell m}Y_{\ell m}(\theta ,\varphi
)+\sum_{\ell =0}^{\infty }\sum_{m=-\ell }^{-1}a_{\ell m}Y_{\ell m}(\theta ,\varphi ) \\
& =\sum_{\ell =0}^{\infty }\sum_{m=0}^{\ell }a_{\ell m}Y_{\ell
m}(\theta ,\varphi
)+\sum_{\ell=0}^{\infty }\sum_{m=-\ell }^{-1}(-1)^{m}\bar{a}_{\ell ,-m}(-1)^{m}\bar{Y}
_{\ell ,-m}(\theta ,\varphi ) \\
& =\sum_{\ell =0}^{\infty }\sum_{m=0}^{\ell }a_{\ell m}Y_{\ell
m}(\theta ,\varphi )+\sum_{\ell=0}^{\infty }\sum_{m=1}^{\ell
}\bar{a}_{\ell m}\bar{Y}_{\ell m}(\theta ,\varphi )
\\
& =\sum_{\ell =0}^{\infty }\sum_{m=0}^{\ell }a_{\ell m}\sqrt{\frac{2\ell +1}{4\pi }\frac{
(\ell -m)!}{(\ell +m)!}}P_{\ell m}(\cos \theta )e^{im\varphi
}+\sum_{\ell =0}^{\infty
}\sum_{m=1}^{\ell }\bar{a}_{\ell m}\sqrt{\frac{2\ell +1}{4\pi }\frac{(\ell -m)!}{(\ell +m)!}}
P_{\ell m}(\cos \theta )e^{-im\varphi }.
\end{align*}
Focussing on $\varphi =0$, we hence obtain
\begin{align*}
T(\theta ,0)& =\sum_{\ell =0}^{\infty }\sum_{m=0}^{\ell }a_{\ell m}\sqrt{\frac{2\ell +1}{
4\pi }\frac{(\ell -m)!}{(\ell +m)!}}P_{\ell m}(\cos \theta
)+\sum_{\ell =0}^{\infty
}\sum_{m=1}^{\ell }\bar{a}_{\ell m}\sqrt{\frac{2\ell +1}{4\pi }\frac{(\ell -m)!}{(\ell +m)!}}
P_{\ell m}(\cos \theta ) \\
& =\sum_{\ell =0}^{\infty }a_{\ell 0}\sqrt{\frac{2\ell +1}{4\pi
}}P_{\ell 0}(\cos \theta
)+2\sum_{\ell =0}^{\infty }\sum_{m=1}^{\ell } { \Re}a_{\ell m}  \sqrt{\frac{2\ell +1}{
4\pi }\frac{(\ell -m)!}{(\ell +m)!}}P_{\ell m}(\cos \theta ).
\end{align*}
For the quadrupole we have that
\begin{align*}
T_{2}(\theta ,0)&:=a_{20}\sqrt{\frac{5}{4\pi }}P_{20}(\cos \theta
)+2\sum_{m=1}^{2} { \Re}a_{2m} \sqrt{\frac{5}{4\pi }\frac{(2-m)!}{(2+m)!
}}P_{2m}(\cos \theta ) \\
& =\sqrt{\frac{5}{4\pi }}\left[ a_{20}P_{20}(\cos \theta
)+2 { \Re}{a_{21}}\sqrt{\frac{1}{3!}}P_{21}(\cos \theta )+2 { \Re}{a_{22}} \sqrt{
\frac{1}{4!}}P_{22}(\cos \theta )\right]\\
&=\sqrt{\frac{5}{4\pi }}\left[  \frac{ a_{20}}{2} (3 \cos^2 \theta-1) -2 { \Re}{a_{21}}\sqrt{\frac{1}{3!}} 3 \sin \theta \cos \theta+2 { \Re}{a_{22}} \sqrt{
\frac{1}{4!}}3 \sin^2 \theta  \right] \\
&=  \sqrt{\frac{5}{4\pi }}\left[  \frac{ a_{20}}{2} (3 \cos^2 \theta-1) -2 { \Re}{a_{21}}\sqrt{\frac{3}{2}}  \sin \theta \cos \theta+2 { \Re}{a_{22}} \sqrt{
\frac{3}{8}} \sin^2 \theta  \right] \text{ .}
\end{align*}
The standard equality $\mathbb{E}[T_{\ell }^{2}]=(2\ell +1) C_\ell/4\pi
=(2\ell +1)/4\pi $
is easily seen to be verified, indeed
\begin{align*}
\mathbb{E}[T_{2}^{2}(\theta ,0)]& =\frac{5}{4\pi }\mathbb{E}\left[
a_{20}P_{20}(\cos \theta )+2\Re {a_{21}}\sqrt{\frac{1}{3!}}P_{21}(\cos
\theta )+2\Re {a_{22}}\sqrt{\frac{1}{4!}}P_{22}(\cos \theta )\right] ^{2} \\
& =\frac{5}{4\pi }\left[ \mathbb{E}[a_{20}^{2}]P_{20}^{2}(\cos \theta )+4
\mathbb{E}[(\Re {a_{21}})^{2}]\frac{1}{3!}P_{21}^{2}(\cos \theta )+4\mathbb{E
}[(\Re {a_{22}})^{2}]\frac{1}{4!}P_{22}^{2}(\cos \theta )\right] \\
& =\frac{5}{4\pi }\left[ P_{20}^{2}(\cos \theta )+4\frac{1}{2\cdot 3!}
P_{21}^{2}(\cos \theta )+4\frac{1}{2\cdot 4!}P_{22}^{2}(\cos \theta )\right]
\\
& =\frac{5}{4\pi }\left[ \frac{1}{4}(3\cos ^{2}\theta -1)^{2}+4\frac{1}{
2\cdot 3!}9\cos ^{2}\theta \sin ^{2}\theta +4\frac{1}{2\cdot 4!}9\sin
^{4}\theta \right] \\
& =\frac{5}{4\pi }\left[ \frac{1}{4}(3\cos ^{2}\theta -1)^{2}+3\cos^{2}\theta \sin ^{2}\theta +\frac{3}{4} \sin
^{4}\theta \right] \\
&=\frac{5}{4\pi }\text{ .}
\end{align*}
On the other hand, we have
\begin{align}
\mathbb{E}[T_{2}^{4}(\theta ,0)]& =\frac{5^{2}}{4^{2}\pi ^{2}}\mathbb{E}
\left[ a_{20}P_{20}(\cos \theta )+2( { \Re}{a_{21})}\sqrt{\frac{1}{3!}}
P_{21}(\cos \theta )+2( { \Re}{a_{22})}\sqrt{\frac{1}{4!}}P_{22}(\cos
\theta )\right] ^{4} \nonumber\\
& =\frac{5^{2}}{4^{2}\pi ^{2}}\left[ \mathbb{E}[a_{20}^{4}]P_{20}^{4}(\cos
\theta )+2^{4}\mathbb{E}[( { \Re}{a_{21}})^{4}]\left( {\frac{1}{3!}}
\right) ^{2}P_{21}^{4}(\cos \theta )+2^{4}\mathbb{E}[( { \Re}{a_{22}}
)^{4}]\left( {\frac{1}{4!}}\right) ^{2}P_{22}^{4}(\cos \theta )\right. \nonumber\\
& \left. \hspace{0.4cm} +6\mathbb{E}[a_{20}^{2}]P_{20}^{2}(\cos \theta )4\mathbb{E}[( {
 \Re}{a_{21}})^{2}]{\frac{1}{3!}}P_{21}^{2}(\cos \theta )+6\mathbb{E}
[a_{20}^{2}]P_{20}^{2}(\cos \theta )4\mathbb{E}[( { \Re}{a_{22}})^{2}]{
\frac{1}{4!}}P_{22}^{2}(\cos \theta )\right. \nonumber\\
& \left. \hspace{0.4cm} +6\cdot 4\mathbb{E}[( { \Re}{a_{21}})^{2}]{\frac{1}{3!}}
P_{21}^{2}(\cos \theta )4\mathbb{E}[( { \Re}{a_{22}})^{2}]{\frac{1}{4!}}
P_{22}^{2}(\cos \theta )\right] \nonumber\\
& =\frac{5^{2}}{4^{2}\pi ^{2}}\left[ 6P_{20}^{4}(\cos \theta )+2^{4}\frac{3}{
2}\left( {\frac{1}{3!}}\right) ^{2}P_{21}^{4}(\cos \theta )+2^{4}\frac{3}{2}
\left( {\frac{1}{4!}}\right) ^{2}P_{22}^{4}(\cos \theta )\right. \nonumber\\
& \left. \hspace{0.4cm} +6P_{20}^{2}(\cos \theta )4\frac{1}{2}{\frac{1}{3!}}P_{21}^{2}(\cos
\theta )+6P_{20}^{2}(\cos \theta )4\frac{1}{2}{\frac{1}{4!}}P_{22}^{2}(\cos
\theta )\right. \nonumber\\
& \left. \hspace{0.4cm} +6\cdot 4\frac{1}{2}{\frac{1}{3!}}P_{21}^{2}(\cos \theta )4\frac{1
}{2}{\frac{1}{4!}}P_{22}^{2}(\cos \theta )\right]\nonumber\\
& =\frac{5^{2}}{4^{2}\pi ^{2}}\left[ 6P_{20}^{4}(\cos \theta )+\frac{2}{
3}P_{21}^{4}(\cos \theta )+\frac{1}{24}P_{22}^{4}(\cos \theta ) +2P_{20}^{2}(\cos \theta ) P_{21}^{2}(\cos
\theta ) \right. \nonumber\\
& \left. \hspace{0.4cm}  + \frac 1 2 P_{20}^{2}(\cos \theta )P_{22}^{2}(\cos
\theta )
+\frac 1 6 P_{21}^{2}(\cos \theta ) P_{22}^{2}(\cos \theta )\right]\nonumber\\
&= \frac{5^{2}}{4^{2}\pi ^{2}}\left[ \frac{39}{8}-9 \cos^2 \theta+
  \frac{189}{4} \cos^4 \theta-81 \cos^6 \theta+ \frac{351}{8} \cos^8
  \theta \right]\text{ .}
\label{eqn:result1}
\end{align}
Direct substitution of $\cos\theta=1$ shows that
eqn.~(\ref{eqn:eqn1}) is indeed fulfilled at the North Pole.
By an analogous computation, we obtain for the octupole:
\begin{align*}
T_{3}(\theta ,0)&: =a_{30}\sqrt{\frac{7}{4\pi }}P_{30}(\cos \theta
)+2\sum_{m=1}^{3}( { \Re}{a_{3m})}\sqrt{\frac{7}{4\pi }\frac{(3-m)!}{
(3+m)!}}P_{3m}(\cos \theta ) \\
& =\sqrt{\frac{7}{4\pi }}\left[ a_{30}P_{30}(\cos \theta )+2\sum_{m=1}^{3}(
 { \Re}{a_{3m})}\sqrt{\frac{(3-m)!}{(3+m)!}}P_{3m}(\cos \theta )\right] \\
& =\sqrt{\frac{7}{4\pi }}\left[ a_{30}P_{30}(\cos \theta )+2( { \Re}{
a_{31})}\sqrt{\frac{2}{4!}}P_{31}(\cos \theta )+2( { \Re}{a_{32})}\sqrt{
\frac{1}{5!}}P_{32}(\cos \theta )+2( { \Re}{a_{33})}\sqrt{\frac{1}{6!}}
P_{33}(\cos \theta )\right] \\
& =\sqrt{\frac{7}{4\pi }} \frac 1 2 \left[ a_{30} (5 \cos^3 \theta-3 \cos \theta)- \sqrt{3}( { \Re}{
a_{31})}  \sin \theta (5 \cos^2 \theta-1)+\sqrt{30}( { \Re}{a_{32})} \cos \theta \sin^2 \theta \right. \\
&\hspace{0.4cm}\left.+\sqrt{5}( { \Re}{a_{33})}  \sin^3 \theta \right] \text{ ,}
\end{align*}
so that, as expected,
\begin{align*}
\mathbb{E}[T_{3}^{2}(\theta ,0)]& ={\frac{7}{4\pi }} \frac 1 4 \left[  (5 \cos^3 \theta-3 \cos \theta)^2+  \frac 3 2 \sin^2 \theta (5 \cos^2 \theta-1)^2+15 \cos^2 \theta \sin^4 \theta +\frac 5 2  \sin^6 \theta \right]\\
&=\frac{7}{4\pi }\text{ ,}
\end{align*}
and we have
\begin{align}
\mathbb{E}[T_{3}^{4}(\theta ,0)]& =\frac{7^{2}}{4^{2}\pi ^{2}}\mathbb{E}\left[ a_{30}P_{30}(\cos \theta )+2(
 { \Re}{a_{31})}\sqrt{\frac{2}{4!}}P_{31}(\cos \theta )+2( { \Re}{a_{32})
}\sqrt{\frac{1}{5!}}P_{32}(\cos \theta ) \right. \nonumber\\
&\left. \hspace{0.4cm}+2( { \Re}{a_{33})}\sqrt{\frac{1}{
6!}}P_{33}(\cos \theta )\right] ^{4} \nonumber\\
& =\frac{7^{2}}{4^{2}\pi ^{2}}\left[ \mathbb{E}[a_{30}^{4}]P_{30}^{4}(\cos
\theta )+2^{4}\mathbb{E}[( { \Re}{a_{31}})^{4}]{\ \frac{2^{2}}{(4!)^{2}}}
P_{31}^{4}(\cos \theta )+2^{4}\mathbb{E}[( { \Re}{a_{32}})^{4}]{\ \frac{1}{
(5!)^{2}}}P_{32}^{4}(\cos \theta )\right. \nonumber\\
& \left. \hspace{0.4cm} +2^{4}\mathbb{E}[( { \Re}{a_{33}})^{4}]{\ \frac{1}{(6!)^{2}}}
P_{33}^{4}(\cos \theta )+6\mathbb{E}[a_{30}^{2}]P_{30}^{2}(\cos \theta )2^{2}
\mathbb{E}[( { \Re}{a_{31}})^{2}]{\ \frac{2}{(4!)}}P_{31}^{2}(\cos \theta
)\right. \nonumber\\
& \left. \hspace{0.4cm} +6\mathbb{E}[a_{30}^{2}]P_{30}^{2}(\cos \theta )2^{2}\mathbb{E}[(
 { \Re}{a_{32}})^{2}]{\ \frac{1}{5!}}P_{32}^{2}(\cos \theta )+6\mathbb{E}
[a_{30}^{2}]P_{30}^{2}(\cos \theta )2^{2}\mathbb{E}[( { \Re}{a_{33}})^{2}]{
\ \frac{1}{6!}}P_{33}^{2}(\cos \theta )\right. \nonumber\\
& \left. \hspace{0.4cm} +6\cdot 2^{2}\mathbb{E}[( { \Re}{a_{31}})^{2}]{\ \frac{2}{4!}}
P_{31}^{2}(\cos \theta )2^{2}\mathbb{E}[( { \Re}{a_{32}})^{2}]{\ \frac{1}{
5!}}P_{32}^{2}(\cos \theta )\right. \nonumber\\
&\left. \hspace{0.4cm} +6\cdot2^{2}\mathbb{E}[( { \Re}{a_{31}})^{2}]{\
\frac{2}{4!}}P_{31}^{2}(\cos \theta )2^{2}\mathbb{E}[( { \Re}{a_{33}})^{2}]
{\ \frac{1}{6!}}P_{33}^{2}(\cos \theta )\right. \nonumber\\
& \left. \hspace{0.4cm} +6\cdot 2^{2}\mathbb{E}[( { \Re}{a_{32}})^{2}]{\ \frac{1}{5!}}
P_{32}^{2}(\cos \theta )2^{2}\mathbb{E}[( { \Re}{a_{33}})^{2}]{\ \frac{1}{
6!}}P_{33}^{2}(\cos \theta )\right] \nonumber\\
& =\frac{7^{2}}{4^{2}\pi ^{2}}\left[ 6P_{30}^{4}(\cos \theta )+2^{4}\frac{3}{
2}{\ \frac{2^{2}}{(4!)^{2}}}P_{31}^{4}(\cos \theta )+2^{4}\frac{3}{2}{\
\frac{1}{(5!)^{2}}}P_{32}^{4}(\cos \theta )\right. \nonumber\\
& \left. \hspace{0.4cm} +2^{4}\frac{3}{2}{\ \frac{1}{(6!)^{2}}}P_{33}^{4}(\cos \theta
)+6P_{30}^{2}(\cos \theta )2^{2}\frac{1}{2}{\ \frac{2}{(4!)}}P_{31}^{2}(\cos
\theta )\right. \nonumber\\
& \left. \hspace{0.4cm} +6P_{30}^{2}(\cos \theta )2^{2}\frac{1}{2}{\ \frac{1}{5!}}
P_{32}^{2}(\cos \theta )+6P_{30}^{2}(\cos \theta )2^{2}\frac{1}{2}{\ \frac{1
}{6!}}P_{33}^{2}(\cos \theta )\right. \nonumber\\
& \left. \hspace{0.4cm} +6\cdot 2^{2}\frac{1}{2}{\ \frac{2}{4!}}P_{31}^{2}(\cos \theta )2^{2}
\frac{1}{2}{\ \frac{1}{5!}}P_{32}^{2}(\cos \theta )+6\;2^{2}\frac{1}{2}{\
\frac{2}{4!}}P_{31}^{2}(\cos \theta )2^{2}\frac{1}{2}{\ \frac{1}{6!}}
P_{33}^{2}(\cos \theta )\right. \nonumber\\
& \left. \hspace{0.4cm} +6\cdot 2^{2}\frac{1}{2}{\ \frac{1}{5!}}P_{32}^{2}(\cos \theta )2^{2}
\frac{1}{2}{\ \frac{1}{6!}}P_{33}^{2}(\cos \theta )\right]\nonumber\\
& =\frac{7^{2}}{4^{2}\pi ^{2}}\left[ 6P_{30}^{4}(\cos \theta )+ \frac 1 6 P_{31}^{4}(\cos \theta )+\frac{4!}{(5!)^2} P_{32}^{4}(\cos \theta )\right. \nonumber\\
& \left. \hspace{0.4cm} + \frac{4!}{(6!)^2} P_{33}^{4}(\cos \theta
)+P_{30}^{2}(\cos \theta )P_{31}^{2}(\cos
\theta )\right. \nonumber\\
& \left. \hspace{0.4cm} + \frac{1}{10}P_{30}^{2}(\cos \theta )
P_{32}^{2}(\cos \theta )+ \frac{1}{60}P_{30}^{2}(\cos \theta ) P_{33}^{2}(\cos \theta )\right. \nonumber\\
& \left. \hspace{0.4cm} +\frac{1}{60} P_{31}^{2}(\cos \theta ) P_{32}^{2}(\cos \theta )+\frac{2}{6!} P_{31}^{2}(\cos \theta ) P_{33}^{2}(\cos \theta )\right. \nonumber\\
& \left. \hspace{0.4cm} +\frac{1}{5 \cdot6!}  P_{32}^{2}(\cos \theta ) P_{33}^{2}(\cos \theta )\right]\nonumber\\
& =\frac{7^{2}}{4^{2}\pi ^{2}}\left[ \frac{147}{32}-\frac{261}{16}
  \cos^2 \theta+\frac{4977}{32} \cos^4 \theta-\frac{5115}{8} \cos^6
  \theta+\frac{40725}{32} \cos^8 \theta-\frac{19125}{16} \cos^{10}\theta \right.\nonumber\\
& \left. \hspace{0.4cm} + \frac{13575}{32} \cos^{12}\theta \right]
\label{eqn:result2}
\text{.}
\end{align}
Once again, direct substitution of $\cos\theta=1$ shows that
eqn.~(\ref{eqn:eqn1}) is indeed fulfilled at the North Pole. Note that
obtaining a polynomial in $\cos\theta$ of degree 12 is expected, as
$T_{\ell}(\theta , \varphi)$ is a cubic polynomial in $\cos\theta$ when
$\varphi=0$.  The expressions for the quadrupole and the octupole are
plotted in Fig.~\ref{fig:analytic_trispectrum} of
Sect.~\ref{sec:results}.

\end{document}